\documentstyle[12pt,epsf]{article}

\font\sf=cmss10                    %San-Serif 10
\def\del{\partial}
\def\half{\frac{1}{2}}
\def\ra{\rightarrow}

\def\Tr{\rm Tr\ }

\newcommand{\mathbold}[1]{\mbox{\boldmath $\bf#1$}}
\def\mJ{\mathbold{J}}
\def\mA{\mathbold{A}}
\def\mB{\mathbold{B}}
\def\mC{\mathbold{C}}
\def\ma{\mathbold{a}}
\def\mb{\mathbold{b}}
\def\mc{\mathbold{c}}
\def\mX{\mathbold{X}}
\def\mG{\mathbold{G}}
\def\mx{\mathbold{x}}
\def\mZ{\mathbold{Z}}
\def\mz{\mathbold{z}}
\def\malpha{\mathbold{\alpha}}
\def\momega{\mathbold{\omega}}
\def\mdelta{\mathbold{\delta}}
\def\bbbz{{\sf Z\!\!\!Z}}
\def\sl2z{SL(2,\bbbz)}
\def\fracs#1#2{\textstyle\frac #1#2}
\def\ll{\lambda \cdot \lambda}

\def\zz{{\bf z}}
\def\z0{{\bf z_0}}
\def\gg{{\cal G}}

\def\bbbq{{\hbox{Q}\!\!\!\hbox{\rule{0.25mm}{2.88mm}\,}}}
\def\bbbp{\,{\hbox{P}\!\!\!\!\!\hbox{I}\,\,\,}}

\newcommand{\be}{\begin{equation}}
\newcommand{\ee}{\end{equation}}
\newcommand{\bea}{\begin{eqnarray}}
\newcommand{\eea}{\end{eqnarray}}
\newcommand{\nn}{\nonumber}
\newcommand{\bean}{\begin{eqnarray*}}
\newcommand{\eean}{\end{eqnarray*}}
\newcommand{\myref}[1]{(\ref{#1})}
\newcommand{\secref}[1]{sec.~\protect\ref{#1}}
\newcommand{\figref}[1]{Fig.~\protect\ref{#1}}

\newcommand{\unit}{1\!\!1}
\newcommand{\mtimes}{\!\times\!}
\newcommand{\mQ}{\ell}
\newcommand{\genh}{\gg^{enh}}
\newcommand{\Genh}{{\bf G}^{enh}}
\newcommand{\comment}[1]{}
\newcommand{\onefigure}[2]{\begin{figure}[htbp]
         \caption{#2\label{#1}(#1)}
         \end{figure}}
\renewcommand{\onefigure}[2]{\begin{figure}[htbp]
         \begin{center}\leavevmode\epsfbox{#1.eps}\end{center}
         \caption{#2\label{#1}}
         \end{figure}}
\begin{document}

\setlength{\baselineskip}{14pt} % Makes the document double spaced
\setlength{\parskip}{1.35ex}
\setlength{\parindent}{0em}

\noindent

\thispagestyle{empty}
{\flushright{\small MIT-CTP-2808\\hep-th/9812209\\}}

\vspace{.3in}
\begin{center}\Large {\bf 
Uncovering Infinite Symmetries on [p,q] 7-branes:
Kac-Moody Algebras and Beyond} 
%Affine and Indefinite Kac-Moody Symmetries on [p,q]~7-branes:
%The role of $\widehat{\bf E}_{\bf 9}$ and ${\bf E_{10}}$}
\end{center}

\vspace{.1in}
\begin{center}
{\large Oliver DeWolfe, Tam\'as Hauer, Amer Iqbal and Barton Zwiebach}

\vspace{.1in}
{ {\it Center for Theoretical Physics,\\
Laboratory for Nuclear Science,\\
Department of Physics\\
Massachusetts Institute of Technology\\
Cambridge, Massachusetts 02139, U.S.A.}}
\vspace{.2in}

E-mail: {\tt odewolfe,hauer,iqbal@ctp.mit.edu, zwiebach@irene.mit.edu}
\end{center}
\begin{center}December 1998\end{center}

\vspace{0.1in}
\begin{abstract}
\noindent
In a previous paper we explored how conjugacy classes of the modular
group classify the symmetry algebras that arise on type IIB [p,q]
7-branes.  The Kodaira list of finite Lie algebras completely fills
the elliptic %conjugacy 
classes as well as some parabolic %conjugacy
classes. Loop algebras of $E_N$ fill additional parabolic classes, and
exotic finite algebras, hyperbolic extensions of $E_N$ and more
general indefinite Lie algebras fill the hyperbolic %conjugacy
classes. Since they correspond to brane configurations that cannot be
made into strict singularities, these non-Kodaira algebras are
spectrum generating and organize towers of massive BPS states into
representations. The smallest brane configuration with unit monodromy
gives rise to the loop algebra $\widehat E_9$ which plays a central
role in the theory. We elucidate the patterns of enhancement relating
$E_8, E_9, \widehat E_9$ and $E_{10}$.  We examine configurations of
24 7-branes relevant to type IIB compactifications on a two-sphere, or
F-theory on K3.  A particularly symmetric configuration separates the
7-branes into two groups of twelve branes and the massive BPS spectrum
is organized by $E_{10} \oplus E_{10}$. 
\end{abstract}

%\end{document}

\renewcommand{\theequation}{\thesection.\arabic{equation}}
\newcommand{\newsection}[1]{\section{#1}\setcounter{equation}{0}}

\newpage
\newsection{Introduction}

One of the most intriguing aspects of string theory is that the entire
massive spectrum appears to be associated with the spontaneous
breaking of a large and rather mysterious symmetry. This symmetry is
not restored in any familiar vacuum, but it clearly reflects the
underlying structure of the theory. Unbroken symmetries, while no
doubt important, are essentially associated to the massless spectrum,
and represent only a small part of the complete structure. One of our
objectives in this paper is to explain how the simple compactification
of F-theory on an elliptic K3, equivalent to type IIB superstrings on
a two-sphere with 24 7-branes \cite{vafa}, provides a natural setting
where infinite symmetries associated to the massive BPS spectrum of
the theory can be identified explicitly. These infinite symmetries are
completely natural generalizations of the finite exceptional
symmetries $E_6, E_7$ and $E_8$ that are by now well understood in the
7-brane setup \cite{GZ,johansen,mgthbz,dewolfezwiebach}. Indeed, the
main difference lies in that the finite algebras arise from brane
configurations where the branes can be brought together, while this is
not the case for the infinite algebras.  Instead the infinite set of
BPS strings and BPS string junctions stretching between the branes
always represent massive states and essentially correspond to the
generators of the infinite dimensional algebra.

%There is a substantial reason
%why certain 7-branes cannot be brought together, if they did the
%complex structure $\tau$ of the elliptic fiber on the K3 would develop
%a negative imaginary part, alternatively, the type IIB string coupling
%would become imaginary. This is the reason why the associated infinite
%algebras are always spontaneously broken and cannot be restored.

This is another example of how interesting configurations of type~IIB
[p,q] 7-branes are not limited to those where the branes can be
brought together.  In a previous paper \cite{finite}, entire families
of configurations realizing finite $ADE$ algebras were explored,
extending the $D_N$, $E_N$ and Argyres-Douglas $H_N$ series beyond
those cases that exist in the Kodaira classification of K3
singularities.  D3-brane probes in the background of such
non-collapsible configurations realize physically relevant 4D theories
with eight supercharges such as $SU(2)$ Seiberg-Witten theory with
$N_f = 0,1,2,3$ flavors, and some of the theories with $E_N$ global
symmetry which also arise in toroidal compactifications of the 6D
tensionless string theory.

Although only finite Lie algebras are associated to singularities in
the Kodaira classification, in configurations that cannot collapse
many types of algebras arise.  As was found in \cite{dewolfe}, affine
exceptional algebras appear whenever the intersection form of string
junctions realized on a brane configuration reproduces the affine
Cartan matrix.  These brane configurations are no more exotic than the
ones realizing their finite counterparts.  Some new features appear in
these cases, including an imaginary root junction corresponding to a
genus one curve in K3, and a relationship between the asymptotic
$(p,q)$ charge of a junction and a Lie algebraic property, the level
$k$ of the associated representation.  More exotic infinite algebras
also appear. We find, for example, hyperbolic exceptional algebras,
many Lorentzian Kac-Moody algebras, as well as loop extensions of
infinite algebras and other algebras that are not even of Kac-Moody
type.

In \cite{finite}, it was seen how 7-brane configurations realizing the
various finite Lie algebras were classified by the total monodromy
around the branes, along with the additional data of the number of
branes and the possible asymptotic charges on supported junctions.
Since the monodromy is only unique up to global $\sl2z$
transformations, this classification determines a correspondence
between 7-brane Lie algebras and conjugacy classes of $\sl2z$.  These
classes are well-studied in the mathematical literature, and are
divided by their $\sl2z$-invariant trace into the elliptic, parabolic
and hyperbolic classes.  The finite algebras spanned the elliptic
classes, and fell into some of the parabolic classes and hyperbolic
classes of negative trace as well.

A powerful mathematical correspondence exists between conjugacy
classes of $\sl2z$ and equivalence classes of binary quadratic forms.
In \cite{finite}, it was seen how this correspondence determines the
asymptotic charge quadratic form $f(p,q)$, introduced in
\cite{dewolfezwiebach} and used in \cite{DHIZ} to find the global
symmetry representations of various $(p,q)$ dyons in the
Seiberg-Witten and $E_N$ theories.  In addition, it was found that
$f(p,q)$ determines how one algebra enhances to another when an
additional $[p,q]$ 7-brane is added: when $f(p,q) < -1$ only a new
$u(1)$ factor appears, $f(p,q) = -1$ enhances $\gg$ to $\gg \oplus
A_1$, and $-1 < f(p,q) < 1$ indicates the enhancement to another
finite algebra.

The various infinite-dimensional algebras arising on 7-brane
configurations fit into the classification by $\sl2z$ conjugacy
classes, filling many of the parabolic and hyperbolic classes of
positive trace.  These algebras are shown to appear when one enhances
a finite algebra with a $[p,q]$ 7-brane satisfying $f(p,q) \geq 1$.
Affine algebras appear for $f(p,q)=1$, while other Kac-Moody algebras
with indefinite Cartan matrix appear when $f(p,q)>1$. All these
Kac-Moody algebras appear when their Cartan matrices are realized in
the junction intersection form.

Not all of the infinite-dimensional algebras arising on 7-branes are
Kac-Moody, however.  A central example which we study in some detail
is the loop algebra $\widehat{E}_9$, which is a loop enhancement of
the affine algebra $E_9$, the same way that the affine algebras can be
viewed as loop algebras of their finite counterparts.  The algebra
$\widehat{E}_9$ has no Cartan matrix and so cannot be Kac-Moody. The
associated configuration, which we call ${\bf \widehat E_9}$, has unit
monodromy; its twelve branes are in fact the fewest that can realize
such a monodromy.  It follows that factors of ${\bf \widehat{E}_9}$
are invisible to the total monodromy of some configuration.  There are
enhancements of $\widehat{E}_9$ as well, all of which are not
Kac-Moody, such as the loop algebras of the $E_N$ series for $N>9$.
The configuration we are associating to $\widehat{E}_9$ was first
studied in \cite{Imamura}, where it was argued that the collapse of
these branes corresponds to a decompactification limit.

The classification, summarized in Table \ref{maintable}, has a rich
and interesting structure.  The collapsible brane configurations
corresponding to Kodaira singularities occupy the elliptic classes and
some of the parabolic classes.  Configurations associated to
hyperbolic classes of negative trace realize various exotic finite
algebras.  The affine exceptional algebras appear in certain parabolic
classes, as do the non-Kac-Moody loop algebras such as
$\widehat{E}_9$.  The hyperbolic classes of positive trace are filled
with various other infinite-dimensional algebras, including hyperbolic
algebras like $E_{10}$.  Laid on top of everything is a kind of
${\widehat{E}_9}$ periodicity, as a given conjugacy class will also
realize configurations with arbitrary numbers of unit-monodromy
${\bf \widehat{E}_9}$ configurations appended.

We also begin to examine the question of which of these algebras can
arise on a physical compactification of F-Theory on K3.  Just as the
finite $D_N$ and $E_N$ algebras organize the BPS spectra of the field
theories mentioned above, the algebra of the entire brane
configuration on K3 should organize the BPS states of Type IIB string
compactifications. We show how global considerations project out
certain junctions looping around all the 7-branes on the sphere, thus
reducing the rank of the algebra appearing.  A particularly
interesting configuration with the 24 branes split in two groups of
twelve branes makes manifest an $E_{10} \oplus E_{10}$
subalgebra. This splitting is very special in that there is a one to
one correspondence between the root lattice of $E_{10} \oplus E_{10}$
and the K3 homology captured by the junction lattice. 

%%% that arises makes manifest an $E_{10} \oplus E_{10}$ subalgebra.
%%% This description of the K3 surface is very special in the
%%% symmetric splitting and also in that there is a one-to-one
%%% correspondence between the algebra lattice and the homology
%%% lattice of junctions. 

The application of various infinite-dimensional Lie algebras to string
theory is an interesting subject studied by many people. Generalized
Kac-Moody algebras have been developed by Borcherds \cite{Borcherds}
based on vertex algebras of toroidal compactification of bosonic
strings. These algebras, particularly $E_{10}$ \cite{KacMoodyW}, were
also investigated by Gebert and Nicolai \cite{Gebert}.  Moore and
Harvey \cite{MooreHarvey1, MooreHarvey2} studied the algebras of BPS
states, and in particular they identified a role for $E_{10}$ in the
context of threshold corrections for ${\cal N} = 2$ $D=4$ heterotic
compactifications \cite{MooreHarvey1}. Generalized Kac-Moody algebras
also arise in the context of fivebranes \cite{dijkgraaf}. Gritsenko
and Nikulin \cite{Nikulin} developed and studied a variant of the
theory of Lorentzian Kac-Moody algebras involving hyperbolic
generalized Cartan matrices.

Our analysis, however, differs from many of the above in that it does
not involve vertex algebras; the relevant Lie algebras are realized
from intersection of cycles in K3.  The use of configurations giving
infinite-dimensional algebras to the calculation of the spectrum of
probe D3-brane theories is still an open problem which we intend to
address in the future.

In section 2, we review a few important properties of 7-branes, string
junctions and the associated algebras.  Affine algebras on 7-branes
are discussed in section 3, as well as their intersection form and the
imaginary root junction.  In section 4 we introduce the loop algebra
$\widehat{E}_9$ and its associated brane configuration.
Infinite-dimensional algebras of indefinite type are studied in
section 5, these include the exceptional hyperbolic series.  In
section 6 we synthesize this study into a classification of all these
configurations according to their $\sl2z$ conjugacy class, as
presented in Table \ref{maintable}.  Section 7 discusses the
application of these results to two manifolds admitting elliptic
fibration, ${\cal B}_9$ and K3.  Finally, the mathematical techniques
used to calculate the multiplicities of $\sl2z$ conjugacy classes are
detailed in the appendix.

\newsection{Review of 7-branes and Lie Algebras} 
\label{reviewsection}

We shall introduce the infinite-dimensional algebras arising on
7-brane configurations by exploring how they arise as enhancements of
more familiar, finite algebras.  Let us first review the 7-brane
configurations realizing finite algebras, and their supported
junctions.

Consider a configuration $\mG\!\!=\!\!{\bf X}_{{\bf z}_1}\ldots {\bf
X}_{{\bf z}_n}$ of 7-branes ${\bf X_{z_i}}$, where we use a
charge-vector notation $\mz_i \! = \! [p_i,q_i]$ to label the charges of
each brane.  Each 7-brane has an associated monodromy
\begin{eqnarray}
K_{\bf z} &=& \unit + {\bf z}{\bf z}^T S = \left({1\!+\!pq\;\;-\!p^2
\atop \;\;\;q^2 \;\;\;1\!-\!pq}\right) \,,
\label{Kdef}
\end{eqnarray}
and the total monodromy $K$ is the product of the individual monodromy
matrices.  The charges of the 7-branes are only invariant up to global
$\sl2z$ transformations $g$, under which $K_{\bf z} \ra g \, K_{\bf z} \,
g^{-1}$, and the relocation of branch cuts.  Moving the branch cut of
the brane ${\bf X_{z_2}}$ through ${\bf X_{z_1}}$ exchanges their
order in the canonical presentation and modifies the charges $\mz_1$,
as
\begin{eqnarray}
\label{crosstrans}
{\bf X}_{\displaystyle\zz_1}  {\bf X}_{ \displaystyle\zz_2}  &=&
{\bf X}_{\displaystyle\zz_2 }\,
{\bf X}_{\,\displaystyle (\zz_1 + (\zz_1 \times \zz_2)\,
\zz_2\, )} \, , \\
\mbox{where}\hspace{.3in} {\bf z}_1 \mtimes {\bf z}_2 
&\equiv& -{\bf z}_1^TS\, {\bf z}_2 = \mbox{det 
$\left({p_1\;\;p_2 \atop q_1 \;\;\;q_2}\right)$} \,. 
\end{eqnarray}
The physics should not depend on either of these transformations, and
configurations will be considered equivalent if they can be mapped
into one another by global $\sl2z$ and moving branch cuts
\cite{mgthbz}.  The total monodromy changes as $K \ra g K g^{-1}$
under global $\sl2z$, and is invariant under relocating branch cuts.

Consider an algebra $\gg$ of rank $r$ realized on $\mG$.  The
objects charged under $\gg$ are string junctions $\mJ$, which are webs
of $(p,q)$ strings with prongs ending on the 7-branes.  A string
junction carries the charge of a 7-brane if it has a prong on the
brane and/or it crosses the brane's branch cut.  These two
possibilities, combined in a way that is invariant under a
Hanany-Witten transformation, define the invariant charge $Q_{{\bf
X}_i}(\mJ)$ of the junction $\mJ$ associated to the 7-brane
$\mX_i$. The set of invariant charges completely determines the
junction's algebraic properties.  The $Q_{{\bf X}_i}$ are integers,
and thus the space of junctions is a lattice.  Instead of the $Q_{{\bf
X}_i}$, a junction on $\mG$ can be characterized by a $\gg$ weight
vector $\lambda$, as well as asymptotic charges $(p,q)$.  The
self-intersection of a junction is its norm on the lattice of
junctions, and is determined by the length-squared of its Lie algebra
weight vector $\ll$ and a binary quadratic form in the asymptotic
charges, $f_K(p,q)$ \cite{dewolfezwiebach}:
\begin{eqnarray}
\label{intersection}
(\mJ, \mJ) = -\ll + f_K(p,q) \,.
\end{eqnarray}
It was shown in \cite{finite} that the asymptotic charge quadratic
form $f_K(p,q)$ associated to $\mG$ is determined simply by the
monodromy\footnote{The correspondence between conjugacy classes of
$\sl2z$ and equivalence classes of binary quadratic forms was
discussed in another context in \cite{Moore}.},
\begin{eqnarray}
\label{ffquad}
f_K(\mz) = \frac{1}{2-t}\ \mz^T S \, K \, \mz
    \equiv \frac{Q_K(\mz)}{2-t} \,,
\end{eqnarray}
where $S = \left({0\,-\!1\atop\!1\;\;0}\right)$ is the usual $\sl2z$
generator and $t \equiv {\Tr} K$.

Although the $\sl2z$ conjugacy class of $K$ carries most of the
information about the algebra, it is not always sufficient for the
classification of the configurations. One has to specify the number of
7-branes, and possible constraints on the $(p,q)$ asymptotic charges
the branes can support. 
%%%t The $\sl2z$-invariant integer $\mQ$ introduced
%%%t in \cite{finite} to characterize the latter was defined in terms of
%%%t the quadratic form $Q_K(\mz)$. As equation \myref{ffquad} suggests,
%%%t however, the definition of $Q_K(\mz)$ becomes problematic when
%%%t $t=2$, which occurs for affine and loop algebras, as we shall
%%%t see.  Therefore in the present work we use a slightly different
%%%t quantity to characterize the constraints on the asymptotic
%%%t charges. 
The latter can be characterized in the following way
\cite{finite}. For the charges $\mz_i$ of all 7-branes in $\mG$,
define

\begin{equation}
\label{l}
\mQ \equiv \left\{\begin{array}{ll}
\gcd\{{\bf z}_i\times{\bf z}_j, \mbox{for all $i,j$}\;
%\mbox{for all}\;\;\; {\bf z}_i\neq{\bf z}_j 
\}\,  &
\mbox{for mutually nonlocal branes}
\\
0 &
\mbox{for mutually local branes}.
\end{array}\right.
\end{equation}

It is manifest that $\ell$ is invariant under global $\sl2z$
transformations, and it can be shown it also does not change when
branch cuts are relocated.  The asymptotic charges $\zz = (p,q)$ are
constrained to obey
\begin{eqnarray}
\zz \times \,  \zz_i = 0 \quad (\mbox{mod}\ \ell) \,,
\label{chargeconstraint}
\end{eqnarray}
where $\zz_i$ denotes the charges of some 7-brane in $\mG$.  The
constraint (\ref{chargeconstraint}) will be the same regardless of the
choice of $\zz_i$ thanks to (\ref{l}).

%%% Without $g$ some linear combination of $p$ and $q$ is constrained
%%% and for any larger configuration, we extract the greatest common
%%% divisor of these linear combinations, and thus quantify the
%%% constraint on the asymptotic charges.

In the next few sections, we shall begin consider a finite
configuration $\mG$ with algebra $\gg$ and proceed to add a single new
brane $\mZ$ characterized by charges $\mz = [p,q]$, exploring the
infinite algebras $\genh$ that can appear on the total configuration
$\Genh = \mG \mZ$.  This procedure was considered for finite
enhancement in \cite{finite} (sec.5), where it was shown that on the
total configuration $\Genh$, a generic junction with zero
asymptotic charges can be expressed as
\begin{eqnarray}
\label{junctioninbasis}
\mJ = -n\ \mz + n\,( p\,\momega^p + q\, \momega^q) + \sum_i a_i \,
\momega^i  \,,
\end{eqnarray}
and can thus be specified by its $\gg$ weight and the invariant charge
$-n$ on the $\mZ$ brane.  Each such junction $\mJ = (n, \lambda)$ is
identified with a root of $\genh$.  The self-intersection $\mJ^2$ of
these junctions satisfies the relation,
\begin{eqnarray}
\ll = -\mJ^2 + n^2 (f_K(\mz) - 1) \,,
\label{ll}
\end{eqnarray}
where $f_K(\mz)$ is the value of the charge quadratic form of $\mG$
evaluated on the charges of $\mZ$.  Since $\ll \geq 0$ and the BPS
condition requires $\mJ^2 \geq -2$ \cite{DHIZ}, equation (\ref{ll})
need not have solutions for all ``grades'' $n$.  Indeed, for the
finite cases considered in \cite{finite}, there are only solutions at
finitely many $n$.  For $f(\mz) < -1$ the only solution is at $n=0$,
meaning that the only roots are the original ones of $\gg$, and $\genh
= \gg \oplus u(1)$.  This $u(1)$ arises since the space of junctions
without asymptotic charges increases its dimension by one.  For
$f(\mz) = -1$ precisely two new roots $(\pm 1,\lambda\!=\!0)$ appear,
giving $\genh = \gg \oplus A_1$ (unless $n=\pm1$ are incompatible with
\myref{junctioninbasis}).  For values $-1 < f(\mz) < 1$ there is a
solution for finitely many nonzero grades, and $\genh$ is some finite
algebra of rank $r+1$.

As we shall see, when $f(\mz) \geq 1$ we find that $\Genh$ realizes
some infinite-dimensional Lie algebra.  In most cases the algebra
$\genh$ can be identified by finding a set of simple roots.  The
simple roots then determine the Cartan matrix and Dynkin diagram of
$\genh$.  Simple roots of Kac-Moody Lie algebras are always real,
meaning the corresponding junctions have $\mJ^2 = -2$.  In most cases
a single new simple root is necessary, and as explained in
\cite{finite}, this enhanced simple root will be $\malpha_0 = (n_0,
\theta_<)$ where the grade $n_0$ is the lowest positive grade
satisfying (\ref{ll}), and $\theta_<$ is the lowest weight of the Weyl
orbit at this grade.  In a few exotic cases two additional (linearly
dependent) simple roots will be necessary to produce a basis.

The results we shall find for how $f(\mz)$ controls the kind of
enhanced algebra $\genh$ are summarized in Table~\ref{enhancef}.  
%%%t
The case $\mG = {\bf A_N}$, where the configuration is composed of
$N+1$ mutually local branes realizing $\gg =A_N$ is exceptional, 
%%%t
adding a mutually local brane enhances to ${\bf A_{N+1}}$, but any
mutually nonlocal brane does not enhance $\gg$ at all, but instead
allows both asymptotic charges to be realized on junctions.

\begin{table}
\begin{center}
\begin{tabular}{|c|c|} \hline
$\rule{0mm}{5mm}f(\mz)$ & $\genh$ \\ \hline
 $f(\mz) < -1$ & $\gg \oplus u(1)$ \\
 $f(\mz) =-1$ & $\gg \oplus A_1$ \\
$-1 < f(\mz) < 1$ & finite \\
$f(\mz) =1$ & affine or loop \\
$f(\mz) >1$ & indefinite \\ \hline
\end{tabular}
\end{center}
\caption{Algebraic enhancements of brane configurations.  Adding one
brane to an existing configuration $\mG$ realizing a finite algebra
$\gg$, the asymptotic charge form $f(\mz)$ of ${\bf G}$ evaluated on
the new brane determines the type of $\genh$. \label{enhancef}}
\end{table}

%%%%%%%%%%%%%%%%%%%%%%%%%%%%%%%%%%%%%%%%%%%%%%%%%%%%%%%%%%%%%%%%%%
\newsection{Affine Enhancement: $f(\mz) = 1$}
\label{affinize}

Consider the case when the charge quadratic form associated to a
finite algebra $\gg$ can take the value $f(\mz) =1$ for some charge
$\mz$. Add a brane with that charge and examine the resulting
enhancement.  In this case equation (\ref{ll}) for the $\gg$ weight
vector of the new roots reduces to
\begin{eqnarray}
\ll = -\mJ^2 \,,
\label{affinell}
\end{eqnarray}
where the dependence on the grade $n$ drops out.  Hence, unlike the
case for $f(\mz) < 1$, this equation can be satisfied for all
$n$. There are solutions with $\mJ^2 = -2$ as long as the $\gg$ weight
vector $\lambda$ is a root, and so there is an infinite tower of root
vectors, with the roots of $\gg$ repeated at each grade $n$.  (There
is an important exception when $\gg$ has, in addition to roots, other
weight vectors satisfying $\ll=2$. In this case the enhanced algebra
is still affine, but it is not $\widehat{\gg}$.  We discuss this in
section \ref{eaffine}.)  Moreover, we have solutions satisfying $\mJ^2
= 0$ with $\lambda=0$, corresponding to an imaginary root junction, as
we describe in the next subsection.  This is precisely the structure
of the affine Lie algebra $ \widehat{\gg}$.  The realization of these
algebras on 7-branes was first studied in \cite{dewolfe}.  Here we
review those results, and simplify and clarify a few issues.  For
background on infinite-dimensional Lie algebras, see \cite{kac}.

\subsection{The imaginary root junction $\mdelta$}

Due to the degeneracy of the affine Cartan matrix, affine algebras
possess an ``imaginary'' root $\delta = \alpha_0 + \sum_{i=1}^r c^i \,
\alpha_i$, where $r$ is the rank of the associated finite algebra and
$c^i$ are the marks of that algebra.  It has zero intersection with
all roots:
\begin{eqnarray}
\label{delpro}
\delta \cdot \alpha_i = 0 \,, \quad i = 0 \ldots r \,,
\end{eqnarray}
and as a consequence it also obeys $\delta \cdot \delta= 0$.  All
vectors of the form $n \, \delta$ for $n \in \bbbz$ are roots of the
affine algebra, with degeneracy equal to $r$.  For finite algebras,
all root junctions satisfy $(\malpha, \malpha) = -2$; in contrast, for
the affine case we expect to see an imaginary root junction $\mdelta$
satisfying $(\mdelta,\mdelta) = 0$.  Because of (\ref{delpro}) we also
must have $(\mdelta,\malpha_i) = 0$ for all $\malpha_i$, and therefore
the imaginary root has $\lambda = 0$. It is identified as the solution
of (\ref{affinell}) with $\mJ^2 = \ll = 0$, and $n=1$.

For the affine configurations constructed in \cite{dewolfe}, it
was noted that $\mdelta$ could always be presented as a string looping
around the brane configuration.  We now explain why this is guaranteed
by the algebraic properties of the monodromy matrices.  Consider the
total monodromy $K_{\mz} \, K$ obtained by adding to a brane
configuration with monodromy $K$ a $\mZ$ brane of charge $\mz$
satisfying $f_K(\mz) =1$. We claim that this total monodromy admits an
eigenvector $\z0 = (p_0,q_0)$ with eigenvalue one:
\be K_{\mz} \, K \z0 = \z0 \,.
\label{loop1}
\ee
Consider a loop of string with charge $\z0 = (p_0,q_0)$ that first
crosses the branch cut of the $\gg$ configuration, becoming $K \z0$,
and then crosses the cut of $\mZ$, becoming $\z0$ again.  Charge
conservation requires that the charge acquired from the $\gg$
configuration is all lost again on the $\mZ$ brane, and hence $K \z0 -
\z0 = n \mz$ for some $n$. Using $(K-\unit)^{-1} =
(K^{-1}-\unit)/(2 - t)$, we
can now calculate $\z0$ in terms of $\mz$, finding
\begin{eqnarray}
\z0 = { n \over 2-t}\,  (K^{-1}-\unit) \; \mz \, ,
\label{loop2}
\end{eqnarray}
where $t = {\Tr} K$ and $n$ can be chosen to be the minimum integer
that permits integer values for $\z0$. We must now verify that
(\ref{loop1}) holds.  Substituting (\ref{loop2}) and recalling the
expression for the monodromy of a single brane (\ref{Kdef}),
(\ref{loop1}) is equivalent to
\begin{eqnarray}
(\unit + \mz \mz^T S) (\unit-K) \, \mz = (K^{-1} -\unit) \, \mz \,.
\end{eqnarray}
Using $K^{-1} = t \unit - K$ and $\mz^T S \mz = 0$, we finally obtain
the condition
\begin{eqnarray}
(\mz^T S K \mz) \, \mz = (2-t) \, \mz \,,
\end{eqnarray}
which by (\ref{ffquad}) holds precisely when $f_K(\mz) = 1$. This
completes the proof: the junction ${\mathbold \delta}$ is a loop with
charge ${\bf z_0}$ as shown in \figref{delta}.
\onefigure{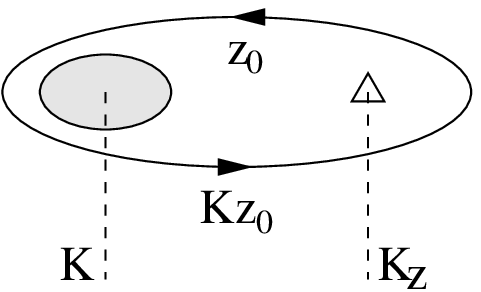}{The imaginary root junction as a loop surrounding the
enhanced configuration.}

Because it traces out a simple Jordan closed curve around the branes,
the junction clearly has $\mdelta^2 = 0$. In addition, it can be chosen
not to intersect any simple root, all of which are localized on the
brane configuration. Therefore all intersection numbers with roots
vanish and the associated weight vector vanishes.  Thus the
presentation of $\mdelta$ as a loop makes its key properties manifest.

\subsection{Configurations realizing affine algebras}
\label{eaffine}

\begin{table}
\begin{center}
\begin{tabular}{|c|l|c|c|} \hline
 $\mathcal{G}$ & Brane Configuration & $K$ & $f_{K}(p,q)$ \\ \hline
\hline & $\rule{0mm}{7mm}$  ${\bf A_N}=\mA^{N+1}$
& $\pmatrix{1&-N-1\cr 0&1}$
&$-\frac{1}{N+1}\, p^{2}$ \\ \cline{2-4}

$\rule{0mm}{7mm}$  $A_N$ &${\bf H_N}={\bf A}^{N+1}{\bf C}$
& $\pmatrix{2&-3-2N\cr 1&-N-1}$
&$\frac{1}{N+1}\{-p^{2}+(N+3)pq-(3+2N)q^{2}\}$ \\ \cline{2-4}

& $\rule{0mm}{7mm}{\bf \tilde{H}_N}={\bf A}^{N}{\bf X}_{\bf [0,-1]}{\bf C}$
& $\pmatrix{1&-N-1\cr 1&-N}$
&$\frac{1}{N+1}\{-p^{2}+(N+1)pq-(N+1)q^{2}\}$   \\ \hline

$\rule{0mm}{7mm}$$D_{N}$  & ${\bf D_N}={\bf A}^{N}{\bf BC}$
& $\pmatrix{-1&N-4\cr0 &-1}$
& $\frac{N-4}{4}\,q^{2}$  \\ \hline

$\rule{0mm}{7mm}$$E_{N}$  & $\rule{0mm}{7mm}$ ${\bf E_N}=
{\bf A}^{N-1}{\bf BCC}~$
& $\pmatrix{-2 & 2N-9\cr -1 &N-5 }$ &
$\frac{1}{9-N}\{p^{2}+(3-N)pq+(2N-9)q^{2}\}$ \\ \cline{2-4}
   &$\rule{0mm}{7mm}$ ${\bf \tilde{E}_N}={\bf A}^{N}
             {\bf X}_{\bf [2,-1]}{\bf C}$
& $\pmatrix{-3&3N-11 \cr -1 &N-4}$
& $\frac{1}{9-N}\{p^{2}+(1-N)pq+(3N-11)q^{2}\}$ \\ \hline
\end{tabular}
\end{center}
\caption{Brane configurations, monodromies and the charge quadratic
form for $A_{N}$, $D_{N}$ and $E_{N}$ algebras. The two series (${\bf
H_N}$ and ${\bf \tilde{H}_N}$) realizing the $A_N$ algebras are
equivalent. The $\mA_N$ series of mutually local branes supports only
asymptotic $p$-charge.
The two series (${\bf E_N}$ and ${\bf \tilde{E}_N}$)
realizing $E_N$ are equivalent for $N\ge2$. \label{series}}
\end{table}

Finite $ADE$ algebras can be realized on 7-branes, as reviewed in
\cite{finite}.  In this section we explore which of the affinizations
of these are realized on branes.  In Table \ref{series}, we summarize
several useful series of brane configurations realizing finite
algebras, and the monodromies and asymptotic charge forms associated
to these configurations.  ${\bf H_N}$ and ${\bf \tilde{H}_N}$ are
equivalent for all $N$, while ${\bf E_N}$ and ${\bf \tilde{E}_N}$ are
equivalent only for $N \geq 2$.  A $[p,q]$ 7-brane is denoted
$\mX_{\bf [p,q]}$, while certain useful 7-branes are abbreviated $\mA
\equiv \mX_{\bf [1,0]}$, $\mB \equiv \mX_{\bf [1,-1]}$, and $\mC
\equiv \mX_{\bf [1,1]}$. 
%%%t
\comment{%%%%%%%%%%%%%%%%%%%%%%%%%%%%%%%%%%%%%%%%%%%%%%%%%%%%%%%%
Recall that two brane configurations are
considered equivalent if they are related by an overall $\sl2z$
transformation and relocation of branch cuts, the latter being
generated by pairwise exchanges of 7-branes according to 
%oliver please cite
\begin{equation}
\label{crosstrans}
{\bf X}_{\displaystyle\zz_1}  {\bf X}_{ \displaystyle\zz_2}  =
{\bf X}_{\displaystyle\zz_2 }\,
{\bf X}_{\,\displaystyle (\zz_1 + (\zz_1 \times \zz_2)\,
\zz_2\, )} ,\hspace{.5in} \mbox{with}
\hspace{.2in}
{\bf z}_1 \mtimes {\bf z}_2 \equiv -{\bf z}_1^TS\, {\bf z}_2 \,.
\end{equation}
} %%%%%%%%%%%%%%%%%%%%%%%%  end comment %%%%%%%%%%%%%%%%%%%%%%%%%%

{\bf Affine exceptional algebras.} Only exceptional affine algebras
$\widehat{E}_N$ seem to appear on 7-branes, while generic
$\widehat{A}_N$ and $\widehat{D}_N$ are absent.  To any ${\bf E_N}$
configuration we can add a brane with $\mz = [3,1]$, and since $f(3,1)
=1$ $\forall \, N$ we obtain affine enhancement (see Table
2). Equivalent configurations are obtained by adding a brane with
charges $g \, \mz$ with $g$ in the centralizer of the monodromy,
$gK_{E_N}g^{-1} = K_{E_N}$, as $f_K(g \mz) = f_K(\mz)$.  We thus
define the series \be {\bf \widehat{E}_N} \equiv {\bf E_N}\, {\bf
X_{[3,1]}} = \mA^{N-1} \mB\mC\mC \mX_{\bf [3,1]} = \mA^{N-1}
\mB\mC\mB\mC \, .  \ee The monodromy is readily calculated to be
\begin{eqnarray}
K(\widehat{E}_N) = \pmatrix{1 & 9\!-\!N \cr 0 & 1} \,.
\label{Kaffine}
\end{eqnarray}
The $\mdelta$ junction is a $(-1,0)$ string going counterclockwise,
and as expected $(-1,0)$ is an eigenvector of the monodromy with
eigenvalue plus one. The invariant charges of $\mdelta$ are
\begin{eqnarray}
\mdelta = \mb + \mc_1 + \mc_2 - \mx_{[3,1]} = \mb_1 + \mc_1 - \mb_2 -
\mc_2 \,,
\end{eqnarray}
and so from the point of view of enhancing ${\bf E_N}$ it has $n=1$
(see\myref{junctioninbasis}).  Note $\mdelta$ has no support on the
$\mA$-branes.

Notice that ${\Tr} K({\bf \widehat{E}_N}) = 2$. This must be the case
for any $\sl2z$ matrix having a unit eigenvalue.  Configurations of
mutually local branes also have a monodromy with this property, but
the string that winds around has the same charge as the branes, and is
therefore equivalent to the trivial zero junction.

The ${\bf \tilde{E}_N}$ configurations may be enhanced as well.
Adding a $[4,1]$ brane gives $f_K(4,1) =1$ (see Table 2) resulting in
\begin{eqnarray}
{\bf \widehat{\tilde{E}}_N} = {\bf A}^N{\bf X_{[2,-1]} C X_{[4,1]}}
\,,
\end{eqnarray}
and in fact $K({\bf \widehat{E}_N}) = K({\bf \widehat{\tilde{E}}_N})$.
The imaginary root $\mdelta$ is still a $(-1,0)$ loop, and has
invariant charges
\begin{eqnarray}
\mdelta = \mx_{[2,-1]} + 2 \mc - \mx_{[4,1]} \,.
\end{eqnarray}

${\bf \widehat{E}_N}$ and ${\bf \widehat{\tilde{E}}_N}$ are equivalent
for $N \geq 2$, since the $\sl2z$ conjugation by $K_A$ required to
demonstrate the equivalence of the finite configurations turns the
$[4,1]$ brane into a $[3,1]$ brane (see equation (2.11) in
\cite{finite}.) We shall use the ${\bf \widehat{E}_N}$ presentation
for simplicity.

For $N=0$, however, there is only ${\bf {\tilde{E}}_0}$ and thus we
must consider ${\bf \widehat{\tilde{E}}_0}\!=\!{\bf X_{[2,-1]} C
X_{[4,1]}}$.  The finite configuration ${\bf \tilde{E}_0}$ has no
algebra, and all supported junctions have nonzero $(p,q)$.  In the
affine case, all junctions with vanishing asymptotic charge are
proportional to the imaginary root $\mdelta$.

%%%%%%%%%%%%%%%%%%%%%%%%%%%%%%%%%%%%%%%%%%%%%%%%%%

For $N=1$ we have ${\bf E_1} \not\cong {\bf \tilde{E}_1}$, and their
affinizations are also inequivalent.  The case of ${\bf \widehat{E}_1}
= {\bf BCBC}$ was studied in detail in \cite{dewolfe}. It was seen to
give the algebra $\widehat E_1 = \widehat A_1 = \widehat {su(2)}$. On
the other hand the configuration \be {\bf \widehat{\tilde{E}}_1} =
{\bf A X_{[2,-1]} C X_{[4,1]} = B A X_{[1,-2]} C} \,, \ee does not
support any zero asymptotic charge junctions with $\mJ^2=-2$, but has
a $\widehat{u(1)}$ algebra instead.  The configuration ${\bf
\widehat{\tilde{E}}_1}$ supports all possible asymptotic charges as
${\bf \tilde{E}_1}$ does, while for ${\bf \widehat{E}_1}$, like ${\bf
E_1}$, only charges obeying $p + q = 0$ (mod 2) are permitted.

When a single $\mA$-brane is added, both ${\bf \widehat{E}_1}$ and
${\bf \widehat{\tilde{E}}_1}$ become equivalent to ${\bf
\widehat{E}_2}$.  The algebra $E_2 = A_1 \oplus u(1)$ is not
semisimple, and $E_3 = A_2 \oplus A_1$ is semisimple but not simple.
The associated affine brane configurations realize the algebras
$\widehat{E}_2$ and $\widehat{E}_3$, defined as the quotients of
$\widehat{A}_1 \oplus \widehat{u(1)}$ and $\widehat{A}_2 \oplus
\widehat{A}_1$, respectively, by central elements.  We will discuss
this further in section \ref{affinesemisimple}.  The configurations
${\bf \widehat{E}_N}$, $N = 4,5,6,7,8$ realize the algebras
$\widehat{A}_4$, $\widehat{D}_5$, $\widehat{E}_6$, $\widehat{E}_7$ and
$\widehat{E}_8$.  These configurations and their algebras are
summarized in Table \ref{etable}.

\begin{table}[t]
\begin{center}
\begin{tabular}{|c|c|c|c|c|c|c|c|c|c|} \hline
$\rule{0mm}{6mm}{\bf \widehat{E}_0}$ & ${\bf \widehat{E}_1}$ & ${\bf
\widehat{\tilde{E}}_1}$ & ${\bf \widehat{E}_2}$ & ${\bf
\widehat{E}_3}$ & ${\bf \widehat{E}_4}$ & ${\bf \widehat{E}_5}$ &
${\bf \widehat{E}_6}$ & ${\bf \widehat{E}_7}$ & ${\bf \widehat{E}_8}$
\\ \hline
$\rule{0mm}{6mm}\{\widehat{0} \}$ & $\widehat{A}_1$ & $\widehat{u(1)}$ &
$\widehat{A}_1 \oplus \widehat{u(1)} / \sim$ & $\widehat{A}_2 \oplus
\widehat{A}_1 / \sim$ & $\widehat{A}_4$ & $\widehat{D}_5$ &
$\widehat{E}_6$ & $\widehat{E}_7$ & $\widehat{E}_8$ \\ \hline
\end{tabular}
\end{center}
\caption{The inequivalent affine exceptional configurations for ${N
\leq 8}$ and their algebras.
\label{etable}}
\end{table}

Finally, the ${\bf \widehat E}_N$ configurations with $N\geq 9$
realize algebras $\widehat{E}_N$ which are identified as loop algebras
of $E_N$. Since $E_N$ with $N\geq 9$ is not a finite algebra, the loop
extensions are not Kac-Moody algebras, and do not have Dynkin diagrams
or Cartan matrices. The simplest case of ${\bf \widehat{E}_9}$ is also
the smallest nontrivial configuration with unit monodromy, and will be
discussed in some detail in section \ref{e9hat}.

{\bf The absence of $\widehat{A}_N$ and $\widehat{D}_N$ algebras.}  We
have claimed that the $\widehat{A}_N$ and $\widehat{D}_N$ affine
algebras do not appear on 7-brane configurations.  Let us discuss
attempting to affinize members of the $\mA$, ${\bf H}$ and ${\bf D}$
series.  Examining the quadratic forms in Table \ref{series}, we see
the equation $f(\mz)=1$ has a solution only for the ${\bf H_4,
H_7,H_8,D_5}$ and ${\bf D_8}$ configurations.  Two of those are
already familiar since ${\bf H_4 \cong E_4}$ and ${\bf D_5 \cong
E_5}$.  We claim that the others also enhance to affine exceptional
algebras.

Take for example ${\bf D_8}$, for which $f(p,q) = q^2$ (see Table
\ref{series}).  Here ${\bf D_8 X_{[p,1]}}$ for any $p$ is an affine
enhancement, since $f(p,1) = 1$.  However the algebra is not
$\widehat{D}_8$ but $\widehat{E_8}$.  A quick way to see this is that
${\bf D_N} \to {\bf E_{N+1}}$ by adding a $[p,1]$ brane, as discussed
in section 6 of \cite{finite} and summarized in this paper in Table
\ref{enhancements}.  (For $p=1$ this is manifest from the definitions
of the ${\bf D}$ and ${\bf E}$ series).  Therefore ${\bf D_8} \to {\bf
E_9} \equiv {\bf \widehat{E}_8}$.  We obtain the larger adjoint ${\bf
248}$ of $E_8$ because the $D_8$ weights with $\ll=2$ include not just
the adjoint ${\bf 120}$, but also the spinors ${\bf 128}$, ${\bf
\overline{128}}$, and so additional root vectors appear upon
enhancement.  Choosing $p=0$ for concreteness, one can show using the
conjugacy rules given in section 6 of \cite{dewolfezwiebach} that the
${\bf \overline{128}}$ junctions cannot end on $\mX_{\bf [p,1]}$,
while the ${\bf 128}$ can end on it with $n$ odd and the ${\bf 120}$
with $n$ even. The new simple root must have the lowest value of
$|n|$, so we must take the lowest root of the spinor ${\bf 128}$,
which can have $n=1$.  We then obtain the Cartan matrix of
$\widehat{E}_8$, not $\widehat{D}_8$, which would have appeared had
the simple root been in the ${\bf 120}$.  The combination of the $D_8$
adjoint ${\bf 120}$ at grade $2n$ and spinor ${\bf 128}$ at grade
$2n+1$ for each $n$ gives a complete $E_8$ adjoint ${\bf 248}$.

Similarly, we have ${\bf H_7} \to {\bf \widehat{E}_7}$ by adding a
$[5,1]$ brane, and ${\bf H_8} \to {\bf \widehat{E}_8}$ by adding a
$[4,1]$ brane.  The latter is in accord with the enhancement ${\bf
H_N}$ to ${\bf E_{N+1}}$ noted in Table \ref{enhancements}.  In both
cases the new simple root arises from additional weights of $\ll=2$.

\noindent
{\bf Equivalence of ${\bf \widehat{E}_8}$ and ${\bf E_9}$}.  Let
us finally give an example showing how two different configurations
realizing $\widehat{E}_8 = E_9$ can be mapped into one another by
global $\sl2z$ transformations and relocating branch cuts.  The
configuration ${\bf E_8} = \mA^7 \mB\mC\mC$ can be enhanced both to
${\bf \widehat{E}_8} = \mA^7 \mB\mC\mB\mC$ and ${\bf E_9} = \mA^8
\mB\mC\mC$, and each realizes an $E_9$ algebra, as discussed in
\cite{dewolfe}.  The ${\bf E_9}$ configuration supports an imaginary
root junction $\mdelta$ which is a $(3,1)$ string looping
counterclockwise, with invariant charges
\begin{eqnarray}
\mdelta = - \sum_{i=1}^8 \ma_i + 4\mb + 2 \mc_1 + 2 \mc_2 \,.
\end{eqnarray}
We shall now prove that
these two configurations are indeed equivalent. We find that
\be
K ({\bf \widehat{E}_8}) = T = \pmatrix{1&1\cr 0&1} \stackrel{g}{\cong}
\pmatrix{-2&9\cr -1 & 4} = -T^2 S T^{-4} = K ({\bf E_9})\,, 
\ee
where similarity is implemented by
\be K({\bf \widehat{E}_8}) = g\, K({\bf E_9})\, g^{-1}\,, \quad\hbox{with}\quad
g= \pmatrix { 0 & -1 \cr 1 & -3}\,.
\ee
This monodromy turns ${\bf
E_9}$ into ${\bf E_9}' = (\mX_{\bf [0,1]})^8 \mX_{\bf [1,4]} \,
(\mX_{\bf [1,2]})^2 $, which can be shown to be equivalent to
${\bf \widehat{E}_8}$ after relocating branch cuts:
\begin{eqnarray}
(\mX_{\bf [0,1]})^8 \mX_{\bf [1,4]} \, (\mX_{\bf [1,2]})^2 &=&
(\mX_{\bf [0,1]})^8 \mX_{\bf [1,2]} \,\mA \,\mX_{\bf [1,2]} =
(\mX_{\bf [0,1]})^7 \mC \mX_{\bf [0,1]}\, \mA \mX_{\bf [1,2]}
\nonumber \\ &=& \mC \mA^7 \mX_{\bf [0,1]}\, \mA \mX_{\bf [1,2]} = \mC
\mA^7 \mB \mC \, \mX_{\bf [0,1]} \nonumber \\ &=& \mC \mA^3 (\mA^4
\mB\mC) \,\mX_{\bf [0,1]} = \mC \mA^3 \mX_{\bf [0,1]}\, \mA^4 \mB\mC
\\ &=& \mC \mX_{\bf [3,1]} \, \mA^7 \mB\mC = \mB\mC \mA^7 \mB\mC =
\mA^6\, \mB \mX_{\bf [0,1]} \mC \mB \mC  \nonumber
\\ &=&\mA^7 \,\mB\mC\mB\mC \,.  \nonumber
\end{eqnarray}

\subsection{The affine intersection form}

Here we discuss the junction intersection form for the ${\bf
\widehat{E}_N}$ configuration $(N\leq 8)$, continuing the discussion
of \cite{dewolfe}.\footnote{ Some notation has been modified; the
junction parameter there called $\tilde{n}$ is now called $n$, while
the Lie algebraic quantity there denoted $n$ is now $\overline{n}$.}
We see how the junction self-intersection is related to the length of
the corresponding affine weight vector.  This formula involves the
asymptotic charge form of the finite ${\bf E_N}$ configuration.

The Cartan matrix of an affine algebra $\widehat{\gg}$ is degenerate,
and therefore it cannot be inverted.  The inner product of two affine
weight vectors is not a straightforward generalization of the finite
case.  In fact the weight vector itself is specified not just by the
Dynkin labels $a_i$, but by a grade $\overline{n}$ as well.  Then
inner product of the $\widehat{\gg}$ weight vectors can be shown to be
\begin{eqnarray}
(\lambda_1\cdot \lambda_2)_{\widehat{\gg}} = (\lambda_1 \cdot
\lambda_2)_{\gg} + \overline{n}_1 \, k_2 + k_1 \, \overline{n}_2 \,,
\end{eqnarray}
where the level $k$ is a particular combination of the Dynkin labels
defined by
\begin{eqnarray}
k \equiv (\lambda \cdot \delta) = a_0 + \sum_{i=1}^r \, c^i \, a_i \,,
\end{eqnarray}
with $c^i$ the marks of $\gg$, and $(\lambda_1 \cdot \lambda_2)_{\gg}$
the inner product of the $\gg$ weight vectors defined by the $a_i$, $i
= 1 \ldots r$.

In our junction picture, we enhance an $E_N$ algebra to
$\widehat{E}_N$ by adding a single brane.  This adds only one new
degree of freedom to the possible junctions appearing on the
configuration, so it is not immediately obvious how the two new affine
quantities, $\overline{n}$ and $a_0$ (or $\overline{n}$ or $k$) can be
accounted for.  What happens is that the Dynkin labels and asymptotic
charges are not independent, but instead the level $k$ is a linear
combination of $p$ and $q$.  In the canonical realization of
${\bf \widehat{E}_N}$,
\begin{eqnarray}
\label{levelq}
k = -q = -(\mJ, \mdelta)\,,
\end{eqnarray}
where, as discussed before, $\mdelta$ is a counterclockwise loop of
charge $(-1,0)$ intersecting no root junction and carrying no
asymptotic charge.  As in the finite case, there are two parameters
not specified by the Dynkin labels.  They are not, however, $p$ and
$q$, but rather $p$ and $n$, the latter being the coefficient of
$\mdelta$ in the junction, and coinciding with the grade $n$ found in
enhancing $E_N \ra \widehat{E}_N$.  In a highest weight affine
representation the Lie algebraic grade $\overline n$ is fixed
conventionally to be zero for the highest weight vector. In the
junction picture the natural grade of a junction is given by
$n$. There can be an overall offset between $n$ and $\overline n$.
For all weight vectors $\lambda$ in an affine representation, and all
associated junctions $\mJ (\lambda) $ we set
\be
\label{gradeshift}
n (\mJ) = \overline{n}(\lambda) + n_0\, ,
\ee
where $n_0$ is a constant throughout the representation. Here $n_0$
can be identified as the junction grade of the highest weight vector.

\onefigure{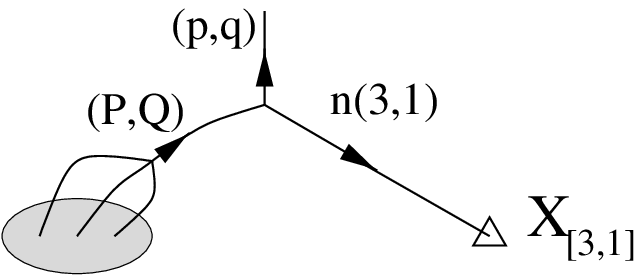}{A junction on an affine configuration.}  Let us now
calculate the intersection form.  As depicted in figure \ref{fig2}, an
arbitrary junction on an ${\bf \widehat{E}_N}$ affine configuration
can be written
\begin{eqnarray}
\mJ = -n\, \mz + P \, \momega^p + Q \, \momega^q + \sum_{i=1}^r \, a_i
\, \momega^i \,,
\end{eqnarray}
with $\mz = [3,1]$ the charges of the enhancing brane.
The total asymptotic charge of $\mJ$ is $(p, q) = -n \,
(3,1) + (P,Q)$. The self-intersection is then
\begin{eqnarray}
\mJ^2 = -n^2 - (\ll)_{E_N} + f_{E_N}(P,Q) + n \, \mz \times
\left( P \atop Q \right) \,,
\end{eqnarray}
where the first term is the contribution of the prongs on $\mX_{\bf
[3,1]}$, the next two are the self-intersection of the ${\bf E_N}$
junction with weight $\lambda$ and charges $(P,Q)$, and the last term
is from the junction point.  In order to express the result in terms
of the true asymptotic charges $(p,q)$ we use
\be
f_{E_N}(P,Q) = f_{E_N} (p + 3n, q + n) = f_{E_N} (p,q) + n (p-q) + n^2
\ee
which follows by explicit computation using the quadratic form listed
in Table \ref{series}. One then arrives at
\begin{eqnarray}
\label{jqaffine}
\mJ^2 = -(\ll)_{E_N} - 2 \, n \, k + f_{E_N}(p,q) \,.
\end{eqnarray}
This is the intersection formula for ${\bf \widehat E_N}$ junctions in
terms of the length squared of the $E_N$ weight vector, the junction
grade $n$ and the level $k$, and the asymptotic charge quadratic form
of ${\bf E_N}$. When $n=0$ we recover the intersection result for
${\bf E_N}$. Since $(\lambda \cdot \lambda)_{\widehat{E}_N} =
(\ll)_{E_N} + 2 \, \overline n \, k $, the first two terms in
(\ref{jqaffine}) are recognized as the affine inner product up to the
shift (\ref{gradeshift}) between $n$ and $\overline n$:
\begin{eqnarray}
\label{jqqaffine}
\mJ^2
   = -(\lambda \cdot \lambda)_{\,\widehat{E}_N} - 2 \, n_0 \, k +
f_{E_N}(p,q) \,,
\end{eqnarray}
The last two terms in the right-hand side are constant over a given
representation.  Thus we see the affine intersection form does not
have the neat partition into Lie algebraic and $(p,q)$ charge parts
present in the finite cases, as in (\ref{intersection}).  Instead the
two are mixed, thanks to the relation (\ref{levelq}) between the level
and the asymptotic charges.

The self-intersection of an affine junction naturally involves the
charge quadratic form of the corresponding finite configuration, but
one may wonder how the quadratic form defined from the affine
monodromy $K({\bf \widehat{E}_N})$ is related to (\ref{jqaffine}).  In
the affine case we have $t=2$, and (\ref{ffquad}) cannot be used..  We
must go back to the original definition of $f_K$ as the
self-intersection of a singlet junction as discussed in section 3 of
\cite{finite}.  For such a singlet only $p$ and $n$ will be nonzero,
since the vanishing of the Dynkin labels forces $q=0$ because of
(\ref{levelq}).  We realize the singlet by a string with charges
$\overline{\mz} = (\overline{p}, \overline{q})$ that crosses the
branch cut of the affine configuration and joins itself to form an
asymptotic string $\mz$. We see $\mz =((9-N)\, \overline{q}, 0)$. This
junction then has intersection
\begin{eqnarray}
\mJ^2 = \mz \times \overline{\mz} = (9-N) \, \overline{q}^{\,2} =
\frac{1}{9-N} \, p^2 \,,
\end{eqnarray}
and defines the affine quadratic form $f_{\widehat{E}_N}$.  This is
just what we expect from (\ref{jqaffine}) for a junction with $a_i = k
= 0$, since $f_{E_N}(p,0) = \frac{1}{9-N}\, p^2$. The fact that $n$
drops out of the intersection formula can be understood since an
asymptotic string with $k=-q = 0$ can wind around the configuration
$n$ times without changing its asymptotic charge or its
self-intersection.  This result is similar to the asymptotic charge
quadratic form for a set of mutually local branes ${\bf A_N}$, where
$f = \frac{1}{N+1}\, p^2$, which should not be surprising since the
monodromies have the same structure: $K({\bf A_N})=
\left({1\;\;-(N+1)\atop\!\!\!\!\!\!\!\!0\;\;\;\;\;\;\;\;\;1}\right).$
We will have more to say about this relationship in section
\ref{classification}.

\subsection{Affine semisimple algebras}
\label{affinesemisimple}

It is not immediately obvious what the affinizations of the non-simple
algebras $E_2 = A_1 \oplus \, u(1)$ and $E_3 = A_2 \oplus A_1$ are,
and the algebras appearing on the configurations ${\bf \widehat{E}_2}$
and ${\bf \widehat{E}_3}$ must be considered carefully.  We examine
${\bf \widehat{E}_3}$ for concreteness, but similar remarks apply to
${\bf \widehat{E}_2}$.

Although one might think the affine enhancement of $E_2$ would be
$\widehat{A_2} \oplus \widehat{A_1}$, this is incorrect.  This algebra
has two distinct imaginary roots, $\delta_1$ and $\delta_2$, and
correspondingly weight vectors have two levels $k_1$ and $k_2$ and two
grades $n_1$ and $n_2$.  The ${\bf \widehat{E}_3}$ configuration,
however, supports only one $\mdelta$ junction, and thus a single level
$k$, as expected since the rank of the resulting algebra
should exceed that of $E_3$ by one.  Furthermore, neither
$\widehat{A_2} \oplus A_1$ nor $A_2 \oplus \widehat{A_1}$ is the
correct algebra, though they have the correct rank.  In these cases,
only one of the simple pieces is affine and has roots at every grade;
the other simple component would have roots only at $n=0$.  Since the
roots of both simple pieces can satisfy (\ref{ll}) at arbitrary grade,
we expect to have the complete root system of $A_2 \oplus A_1$ at
every grade $n$.

\begin{figure}[h]
\begin{center}\leavevmode\epsfbox{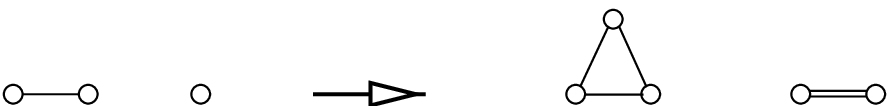}\end{center}
         \caption{The enhancement of the Dynkin diagram of $E_3$ to that
of $\widehat{E}_3$. \label{fig3}}
         \end{figure}

We claim that the resulting algebra $\widehat{E}_3$ is the quotient algebra
$(\widehat{A}_2 \oplus
\widehat{A}_1) / (K_1 - K_2)$, where the ideal in the denominator
is the difference between the two central terms $K_1$ and $K_2$. This retains
only states which have levels satisfying $k_1 = k_2$, ensuring there
is only a single level $k$.  Let $\alpha_1$ and $\alpha_2$ denote the
roots of $A_2$ and $\beta_1$ the root of $A_1$. We then introduce new
simple roots $\alpha_0$ and $\beta_0$, defined as
\be
\alpha_0 = -(\alpha_1 + \alpha_2) + \delta\, , \quad \beta_0 = -\beta_1 +
\delta\,,
\ee
where both new simple roots make use of the unique imaginary root $\delta$.
The inner product $\alpha_0 \cdot \beta_0$ vanishes and the Dynkin diagram of $\widehat E_3$, given in figure \ref{fig3},
coincides with that of $\widehat{A}_2 \oplus \widehat{A}_1$. Note, however,
that
the simple roots of $\widehat E_3$ are not all linearly independent, but
satisfy the constraint
\begin{eqnarray}
\label{constrainte3}
\alpha_0 + \alpha_1 + \alpha_2 = \beta_0 + \beta_1 \,.
\end{eqnarray}
Representations of $\widehat{E_3}$ are products of one representation
each from $A_2$ and $A_1$, constrained to have the same level. 

Similarly, $\widehat{E}_2 = (\widehat{A}_1 \oplus \widehat{u(1)})/(K_1
- K_{u(1)})$. Because of the $u(1)$ factor there is no Dynkin
diagram, and the closest statement analogous to the constraint
(\ref{constrainte3}) is just that the two imaginary roots are
identified by the quotient.

\newsection{The loop algebra ${\bf \widehat{E}_9}$}
\label{e9hat}

The ${\bf \widehat{E}_N}$ sequence of configurations is well-defined
for all $N\geq 0$. We have already considered the cases $N\leq 8$ in
some detail. In this subsection we will examine the configuration
${\bf \widehat{E}_9}$ and will describe concretely the algebra
$\widehat E_9$ it realizes.  The brane configuration $\mA^8
\mB\mC\mB\mC$ we are now identifying with ${\bf \widehat {E}_9}$ was
first considered in \cite{Imamura}, where it was obtained by bringing
together two groups of six branes, each defining a ${\bf D_4}$
configuration. 

As can be seen from (\ref{Kaffine}), the monodromy is $K({\bf
\widehat{E}_9}) = \unit$.  Due to the trivial monodromy, a string of
any charge can wind around the configuration and come back to
itself. Hence ${\bf \widehat{E}_9}$ supports two independent imaginary
root junctions $\mdelta_1$ and $\mdelta_2$, having zero intersection
with all roots including themselves and each other, which can be taken
to have any linearly independent charges.

We have discussed how ${\bf E_8}$ can be enhanced to either of two
equivalent configurations ${\bf E_9} = \mA^8 \mB \mC \mC$ and ${\bf
\widehat{E}_8} = \mA^7 \mB \mC \mC \mX_{\bf [3,1]}$.  As depicted in
figure \ref{fig4}, we can visualize ${\bf \widehat{E}_9}$ as the ${\bf
E_8}$ configuration with both of these enhancements.  The $\mdelta_i$
are then the two imaginary roots associated with the two distinct
affinizations of ${\bf E_8}$.  The first imaginary root $\mdelta_1$
can be taken to be a $(-1,0)$ loop around ${\bf E_8}$ and the $[3,1]$
brane, while the second root $\mdelta_2$ can be taken to be a $(3,1)$
loop around ${\bf E_8}$ and the $[1,0]$ brane, both counterclockwise.

It should be noted that $\mdelta_1$ can be made to pass through the
$\mA$ brane and $\mdelta_2$ can be made to pass through $\mX_{\bf
[3,1]}$, as string segments can pass through a brane of the same
charges without effect.  We can thus deform to a presentation where
both imaginary roots surround the full configuration, making it
manifest that they do not intersect. For any given junction $\mJ$
there will be two levels $k_i = - (\mJ , \mdelta_i)$ and two junction
grades $n_i$ given by the number of prongs on the enhancing $\mA$ and
$\mX_{\bf [3,1]}$ branes.

\onefigure{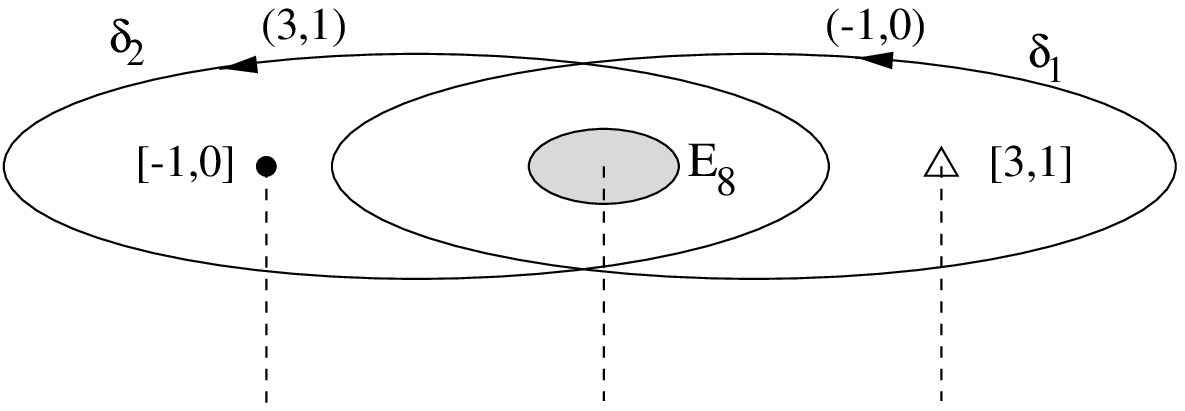}{${\bf \widehat{E}_9}$ viewed as an enhancement of ${\bf
E_8}$ (grey oval) by branes with charges $[-1,0]$ and $[3,1]$.  Both
imaginary roots are presented as loops with certain charges.}

The natural candidate for $\widehat E_9$ is the
double loop algebra of $E_8$ with two central extensions.  Since $E_9$
is the loop algebra of $E_8$, the double loop algebra in question is
the loop algebra of $E_9$. While $E_9$ already has a grade and a
central element, forming the loop algebra of $E_9$ will introduce a
new grade for every generator except the central one, and then a
second nontrivial central extension can be added.  
Hence the
algebra consists of the generators $\{ T^a_{n_1,n_2}, K_1, K_2, D_1,
D_2 \}$ obeying the commutation relations
\begin{eqnarray}
\left[ T^a_{n_1,n_2} , T^b_{m_1,m_2} \right] &=& f^{ab}_{~\;\;c} \,
T^c_{n_1+m_1, 
n_2+m_2} + \kappa^{ab} \left( n_1 \delta_{n_1+m_1} K_1 + n_2
\delta_{n_2+m_2} 
K_2 \right) \,, \nn \\
\left[ D_i, T^a_{n_1,n_2} \right] &=& - n_i \, T^a_{n_1,n_2} \,, \\
\left [K_i, T^a_{n_1,n_2} \right] &=& \left[ K_i, D_j \right] = \left[
D_1, D_2 \right] = 0 \,, \nn
\end{eqnarray}
where $a,b,c$ label the adjoint of $E_8$, $\{ f^{ab}_{~\;\;c} \}$ are
the $E_8$ structure constants, and $\kappa^{ab}$ is the $E_8$ Killing
form.  Just as a loop algebra is the algebra of maps $S^1 \ra \gg$ for
$\gg$ finite and admits a central extension, the above algebra is
algebra of maps $T^2 \ra E_8$ with two central extensions; we have
\begin{eqnarray}
T^a_{n_1,n_2} = T^a \otimes z_1^{n_1} \otimes z_2^{n_2} \,, \quad
D_i = - z_i \frac{\del}{\del z_i} \,,
\end{eqnarray}
with $\{ T^a \}$ the $E_8$ generators and the coordinates $z_1$, $z_2$
parameterizing the two circles.

Note that $\widehat E_9$ is not a Kac-Moody algebra, as it does not
have a Cartan matrix.  The necessary conditions on Cartan matrices of
Kac-Moody algebras correspond to
restrictions on the simple roots, including $(\alpha_i, \alpha_i) = 2$
and $(\alpha_i, \alpha_j) \leq 0, i \neq j$.  The $\delta_i$ are
unacceptable as simple roots since $(\delta_i,\delta_i) = 0$.  An
attempt to find two new simple roots in the spirit of affine algebras
leads to $\alpha_0^{(i)} = \delta_i + \theta_<$, with $\theta_<$
the lowest root of the $E_8$ adjoint.  However we find that
$\alpha_0^{(1)} \cdot \alpha_0^{(2)} = +2$, violating the second
condition.  There is no basis of simple roots simultaneously
satisfying both.
It is a nontrivial fact that the usual loop algebras 
with central extension of finite cases are equivalent to Kac-Moody algebras
with an affine Cartan matrix.  For the double loop
algebra, there is no such correspondence.

We now proceed to examine the intersection form for $\widehat {\bf
E}_9$.  Consider an arbitrary junction $\mJ$ with asymptotic charges $(p,q)$,
as depicted in figure \ref{fig5}. 

\onefigure{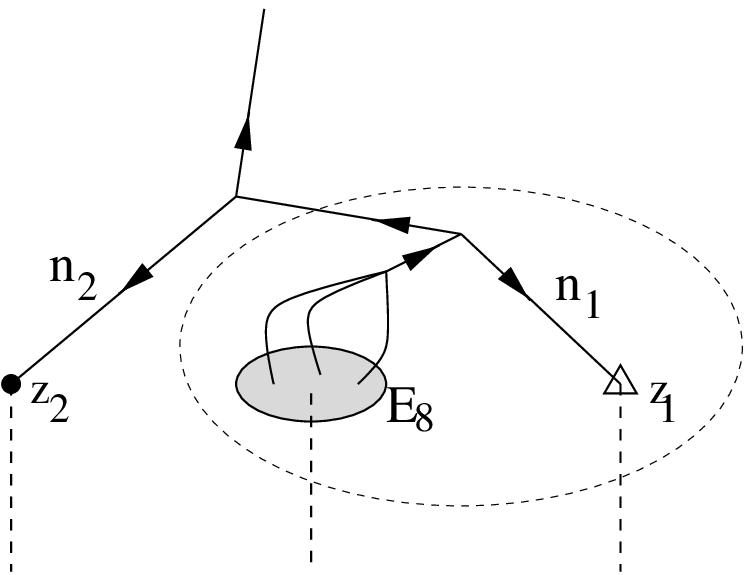}{Calculating the intersection form for $\widehat{E}_9$.}

First consider the sub-junction on the ${\bf \widehat{E}_8}$
sub-configuration, as enclosed by the dashed line.  The
self-intersection of this piece, denoted $(\mJ_9)^2$, is given by
(\ref{jqaffine}) with $n_1$ in place of $n$, and charges $(p+n_2,q)$.
The total self-intersection $\mJ^2$ is then the sum of $(\mJ_9)^2$,
$-(n_2)^2$ for the prongs on the $[1,0]$ brane, and a determinant
giving the contribution at the intersection point:
\begin{eqnarray}
\mJ^2  &=& (\mJ_9)^2 - (n_2)^2 + \hbox{det} 
\pmatrix{p & n_2 \cr q & 0} \,, \nn \\ 
&=& - (\lambda_8 \cdot \lambda_8 + 2n_1 k_1) + f_{E_8} 
(p+n_2\, ,q) - n_2^2 - qn_2 \,.
\end{eqnarray}
The levels of $\mJ$ are given in terms of asymptotic charges by
\begin{equation}
\label{grlev}
k_1 = - q \,, \qquad k_2 = 3q - p \,.
\end{equation}
Solving for $p$ and $q$ in terms of the levels $k_1$ and $k_2$, and
substituting the explicit form of the $E_8$ quadratic form we obtain
\begin{equation}
\label{fine9loop}
\mJ^2 = - \bigl( \lambda_8 \cdot \lambda_8 + 
2n_1 k_1 + 2n_2 k_2\bigr) + (k_1^2 + k_1 k_2 + k_2^2) \,.
\end{equation}

The first group of terms is identified (up to shifts in grades) with
the $\widehat E_9$ length squared of the associated weight vector.
The two levels and grades enter symmetrically in the result. Note that
while for the loop algebras of finite algebras one linear combination
of asymptotic charges is determined by the level, here both asymptotic
charges are determined by specifying the two levels.  Thus ${\bf
\widehat{E}_9}$ is the only configuration we have examined where the
algebraic properties of the associated weight vector completely
specify a junction.  It seems clear that to have double loop algebras
we need configurations of branes with unit monodromy.

There are twelve branes in ${\bf \widehat{E}_9}$, and it is possible
to show that the monodromy of fewer then twelve branes is necessarily
nontrivial\footnote{We are grateful to Don Zagier for demonstrating
this to us.}.  Consider the commutator subgroup $G_C\subset G= \sl2z$
which is the normal subgroup generated by commutators $ghg^{-1}h^{-1},
g,h\in G$.  The quotient $G/G_C$ is abelian and is in fact the cyclic
group ${\bbbz}_{12}$.  $\sl2z$ is generated by two elements $S$ and
$U\ (=ST)$ which satisfy $S^{2}=U^{3}=-\unit$, and then
\begin{equation}
\label{listconj}
\{\unit \,, -U^2S\,,U\,, S\,, U^2\,, US\,, 
-\unit\,,\, U^2S\,, -U\,, -S\,, -U^2\,, -US \,\}.
\end{equation}
is a set of representatives from each coset of $G/G_C$.  It is readily
verified that this is isomorphic to the (additive) cyclic group of
twelve elements, ordered $\{0,1,2, \ldots 11\}$.  Consider now the
canonical homomorphism $\mu:\sl2z \rightarrow \bbbz_{12}$ defined by
$\mu (g) = g G_C$. Note that $\mu(hgh^{-1}) = \mu (g)$, and since the
monodromy of any 7-brane is conjugate to that of a single $\mA$-brane,
$K_A = \left({1 \,-\!1}\atop{\!0\;\;\,1} \right)=-U^2S$, $\mu$ maps it
to $1 \in \bbbz_{12}$, and maps the overall monodromy of $N$ 7-branes
to $N$ (mod 12) $\in \bbbz_{12}$.  As $\mu (\unit) = 0$, only products
of $N = 12k$ 7-brane monodromies for some $k\in \bbbz$ can multiply to
the identity in $\sl2z$.  Only multiples of twelve branes can possess
unit monodromy.

The twelve branes making up the elliptically fibered form of the ninth
del Pezzo surface ${\cal B}_9$, also known as $\half K3$, are
precisely those of $\widehat {\bf E}_9 $.  Two such ${\bf
\widehat{E}_9}$ configurations define a complete elliptically fibered
K3.  The global properties of the bases of these elliptic fibrations
place additional constraints on the lattice of junctions and thus the
algebra, as we shall explore in section \ref{compactifications}.  In
particular, loops around the total configuration of branes on the base
of either ${\cal B}_9$ or K3 are actually trivial since the base is a
sphere.

Note that ${\bf \widehat{E}_N}$ configurations with $N>9$ have the
monodromy of $(N-9)$ $[1,0]$-branes.  This is because the monodromy of
the ${\bf \widehat{E}_9}$ sub-configuration is one, and only the
contribution of the additional $(N-9)$ $[1,0]$-branes is visible. This
does not mean the algebra is elementary. The next term in the series,
for example is ${\bf \widehat{E}_{10}}$ and as its name suggests we
associate it with $\widehat E_{10}$, the loop algebra of
$E_{10}$. While all $E_N$ algebras are Kac-Moody, for $N\geq 9$ their
loop extensions $\widehat E_N$ are not.

%%%%%%%%%%%%%%%%%%%%%%%%%%%%%%%%%%%%%%%%%%%%%%%%%%%%%%%%%%%%%%%%%%%

\newsection{Indefinite enhancement: $f(\mz) > 1$}
\label{indefinite}

We have seen in the previous section that it is possible to realize
infinite dimensional algebras such as $\widehat E_9$ that are not
Kac-Moody. On the other hand, it is also straightforward to find cases
of infinite-dimensional Kac-Moody algebras that are not affine. These
are called indefinite Kac-Moody algebras on account of the indefinite
signature of their Cartan matrix.  We shall see that indefinite
Kac-Moody algebras appear on 7-branes when one adds to a finite
algebra configuration an enhancing brane $\mZ$ with charges satisfying
$f(\mz) >1$.  We will identify explicitly the brane configurations
giving the hyperbolic extension of the exceptional series $E^H_N$,
including $E_{10} = E^H_8$.  The exceptional series $E_N$ for $N>10$
consists of Lorentzian Kac-Moody algebras that are not hyperbolic.  We
also describe a few other indefinite examples, including a class of
algebras with an overcomplete basis of simple roots with a constraint,
and the complete series of strictly hyperbolic algebras of rank 2.

For enhancement to an affine algebra from a simple one, we required
$f(\mz)=1$, and the $n$ dependence of equation (\ref{ll}) dropped out.
As a result, any Weyl orbit satisfying $\ll = -\mJ^2$ appeared at all
grades. On the other hand when $f(\mz)>1$, $f(\mz) - 1$ is positive,
and thus as $|n|$ increases in (\ref{ll}), the length-squared of
acceptable Weyl orbits for a given $\mJ^2$ grows.  Unlike the affine
root systems, which were infinite but repeated the same set of finite
root vectors at all grades, the indefinite root systems grow
uncontrollably, with larger finite weights at each grade.  In
addition, in contrast to the affine case there are clearly solutions
with $\mJ^2 > 0$. In fact, typically for a given positive even $\mJ^2$
there exist infinitely many roots, with arbitrarily large $|n|$ and
$\ll$.

Since indefinite Cartan matrices are invertible, the intersection form
for the associated brane configurations has the same elementary
decomposition into Lie algebra and asymptotic charge blocks as the
finite case, as equation (\ref{intersection}).  Once again $\ll$ is
defined by the inverse Cartan matrix, and $f(p,q)$ is the contribution
to the intersection from asymptotic charges, as in (\ref{ffquad}).
Notice how this is much simpler than the affine case.

\subsection{Hyperbolic exceptional algebras}

The Dynkin diagram of a hyperbolic algebra contains only affine and
finite subdiagrams, and therefore hyperbolic algebras are the simplest
kind of indefinite Kac-Moody algebras.  Given a finite algebra $\gg$
the canonical hyperbolic extension $\gg^H$ is obtained by attaching
two new nodes to the Dynkin diagram: the first node is the one giving
the affine extension $\widehat \gg$, and the second node is attached
to the first and to no other node \cite{kac}.  Perhaps the most
familiar hyperbolic algebra is $E_{10} = E_8^H$, which has as root
lattice the unique even unimodular lattice with Lorentzian $(9,1)$
signature.  As discussed in the introduction, its possible relevance
to string theory has been the subject of much interest.  Hyperbolic
algebras have been studied somewhat by mathematicians but, their root
multiplicities and representation theory remain fairly mysterious.

Let us begin with an example.
Consider ${\bf E_8} =  \mA^7 \mB\mC\mC$, which can be enhanced to various
algebras of rank 9 by adding a single brane.  The charge quadratic
form, as given in table (\ref{series}), is
\begin{eqnarray}
f_{E_8}(p,q) = p^2 - 5pq + 7q^2 \,.
\end{eqnarray}
Charges $\mz = [p,q]$ giving $f(\mz)=1$ enhance to ${\bf E_9}
\cong {\bf \widehat{E}_8}$.  The next smallest possibility is
$f(\mz)=3$, which arises for charges $[1,1]$, among others. The
monodromy matrix $K$ of this configuration satisfies ${\Tr} K
=4$. Since $f(\mz) > 1$, this algebra will be indefinite. We search
for a new simple root, so we focus on junctions with $\mJ^2 = -2$.
Using (\ref{ll}), such junctions $\mJ = (n,\lambda)$ obey
\begin{eqnarray}
\ll= 2 + 2n^2 \,.
\end{eqnarray}
At $n=1$ we must have $\ll=4$, which is satisfied only by the Weyl
orbit with highest weight $\omega^1$. This Weyl orbit contains 2160
weight vectors, and is the largest Weyl orbit of the {\bf 3875}.  We
pick the new simple root $\malpha_0 = (1,-\omega^1)$, which attaches
an extra node to $E_8$ 
%%%t Dynkin diagram as shown in the figure,
giving us the Dynkin diagram $T_{5,4,2}$, associated to the hyperbolic
algebra $E_7^H$.  In the table below we give the length-squared $\ll$
for the Weyl orbits of the real roots and the first few sets of
imaginary roots, for the lowest grades.

\begin{center}
\begin{tabular}{|c|c|c|c|c|} \hline
$\rule{0mm}{5mm}$ $\ll$& $\mJ^2 = -2$ & $\mJ^2 = 0$ & $\mJ^2 = 2$ &
$\mJ^2 = 4$\\ \hline
$|n| = 0$ & $2$ & $0$ & -- & --\\
$|n| = 1$ & $4$ & $2$ & $0$ & --\\
$|n| = 2$ & $10$ & $8$ & $6$ & $4$ \\
$|n| = 3$ & $20$ & $18$ & $16$ & $14$ \\ \hline
\end{tabular}
\end{center}

We could have obtained $E_7^H$ by enhancing $\widehat{E}_7$ as
well. Indeed, starting with ${\bf E_7} = \mA^6 \mB\mC\mC$ we enhance
to ${\bf \widehat{E}_7} = \mA^6 \mB\mC\mC \mX_{\bf [3,1]}$ and get the
hyperbolic configuration by adding yet another $[3,1]$ brane: ${\bf
E^H_7} = \mA^6 \mB\mC\mC ( \mX_{\bf [3,1]})^2$.  The first $[3,1]$
brane creates a node for the root junction $\malpha_0$, a junction
that has a single prong on that brane.  Once we add the second $[3,1]$
brane we obtain a new junction joining the two $[3,1]$ branes,
representing a new simple root $\malpha_{-1}$ producing a new node
that only attaches to the first added node. Therefore in general we
have the configurations
\be
\label{ehyperbolic}
{\bf E^H_N} = \mA^{N-1} \mB\mC\mC (\mX_{\bf [3,1]})^2 \,,
\ee
realizing the hyperbolic Lie algebra $E_N^H$ for $N \leq 8$, and
``hyperbolic'' enhancements of the non-Kac-Moody loop algebras
discussed in section \ref{e9hat} for $N > 8$.  The monodromy is
\begin{eqnarray}
\label{hypK}
K({\bf E_H^N}) = \pmatrix{4 & 27-4N \cr 1 & 7-N} \,.
\end{eqnarray}
All these
configurations for $N \leq 7$ can also be obtained by adding a single
brane to a configuration with a finite Lie algebra of rank $N+1$, as
we saw above for the case of ${\bf E_8} \to {\bf E_7^H}$.

\onefigure{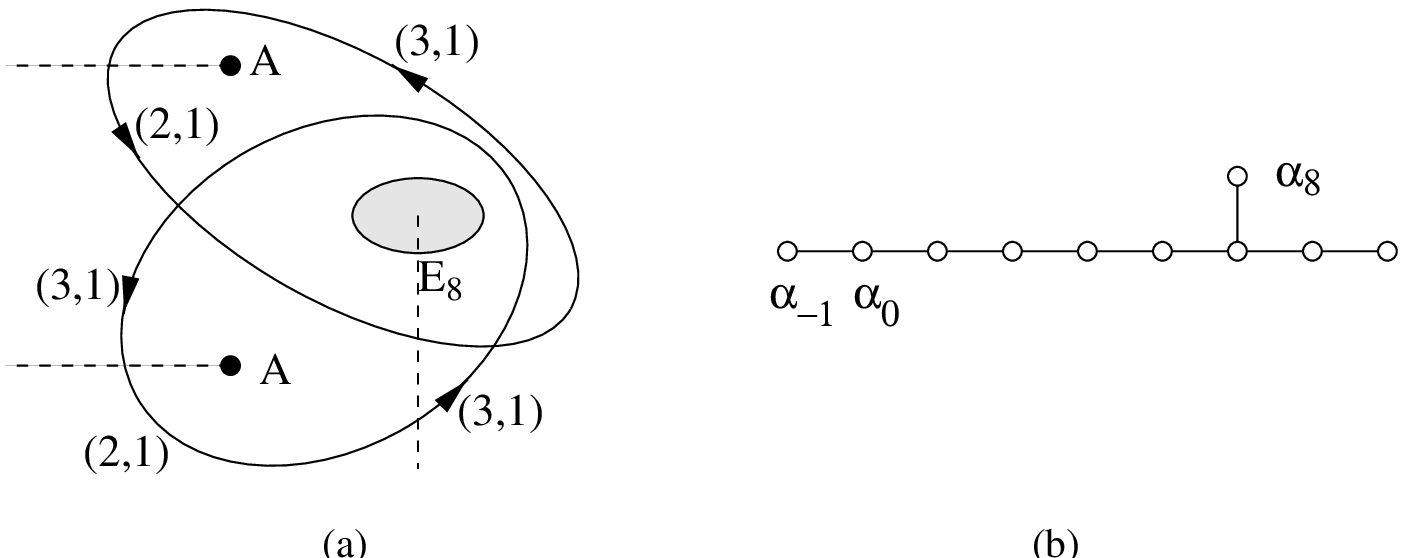}{(a) The ${\bf E_{10}}$ configuration viewed as an
enhancement of ${\bf E_8}$ (grey oval) by two $[1,0]$ branes, with
the two intersecting imaginary roots indicated. The only nonzero
contribution to the intersection comes from where the $(2,1)$ string
segment crosses the $(3,1)$ one. (b) The $E_{10}$ Dynkin diagram.}

The configuration ${\bf E^H_8}$ as defined in (\ref{ehyperbolic})
realizes the algebra $E_8^H = E_{10}$, and is equivalent to a member
of the ${\bf E_N}$ series, ${\bf E_{10}} = \mA^9 \mB \mC \mC$.
We use this presentation, illustrated in figure \ref{fig6},
to highlight the difference between the algebras $E_{10}$ and
$\widehat{E}_9$.  The latter, as explained in section \ref{e9hat},
can be thought of as $E_8$ with two additional imaginary roots
which have zero intersection with all roots, including each other.
On the other hand, $E_{10}$ can be thought of as $E_8$ with
two new roots $u_i$ that are ``imaginary'' in the sense that they
have $u_i \cdot u_i = 0$ and $(u_i, \alpha) = 0$ for all roots $\alpha$
of $E_8$, but which also have nonzero intersection with each other:
$u_1 \cdot u_2 = 1$.  Then the affine root is $\alpha_0 = -u_1 +
\theta$, and the hyperbolic root is $\alpha_{-1} = u_1 + u_2$
\cite{kac}.

The brane picture makes this manifest.  Either one of the two enhancing
$\mA$ branes alone would produce ${\bf E_9}$ with an imaginary root; these
two $\mdelta_i$ correspond to the two $u_i$.
%oliver  maybe call the junctions delta_1 and \delta_2 in the e10 figure
% in analogy to the \hat e9 figure (and the above text)
When both enhancing branes are present, we still have both imaginary
roots.  Note that the loops intersect nontrivially; at one of their
crossing points their charges coincide but at the other one they do
not.  Having nontrivial intersection, it is impossible to have any of
these loops encircle the whole configuration. Indeed the total
monodromy does not allow such loop junction, consistent with $E_{10}$
not being a loop algebra.

\subsection{Algebras with two enhanced simple roots}

Another notable set of
enhancements of exceptional finite algebras is characterized by more
than one Weyl orbit having the value of $\ll$ necessary to solve
(\ref{ll}) at $n=1$.  It is then not possible to choose a single new
simple root that spans the entire set of new roots, since if it is
chosen from one of the Weyl orbits some roots from the other will be
missed.  These indefinite algebras have an interpretation: they are
the enhancements of affine semisimple algebras.

\onefigure{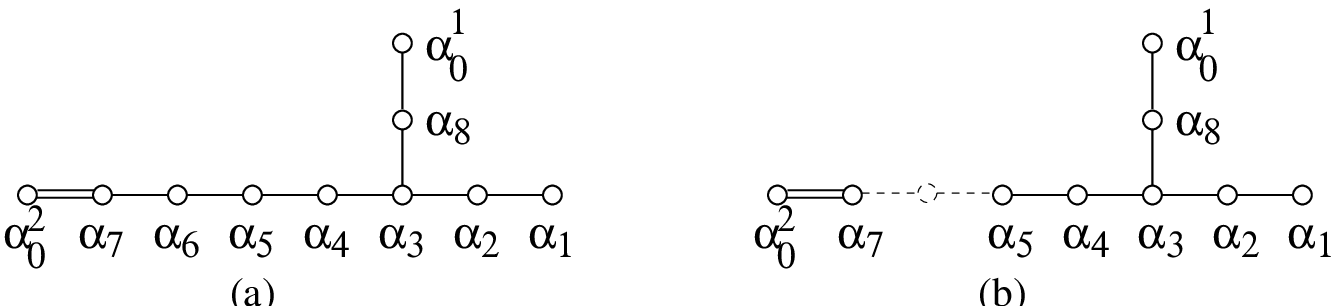}{(a) The Dynkin diagram resulting from adding both
new simple roots to $E_8$. (b) With the node corresponding to
$\alpha_6$ removed, Figure \ref{tworoots} becomes the Dynkin diagram
of the affine version of $E_6 \oplus A_1$. }

We give an example, again with ${\bf E_8}$.  Adding a $[0,1]$ brane
gives $f(0,1) = 7$, the next smallest value after $f=3$.  Then at
$n=1$ we can have new real roots with $\ll = 8$.  There are two $E_8$
Weyl orbits with this length-squared, with highest weight vectors
$\omega^8$ and $2 \omega^7$.  We cannot choose a single simple root,
but instead have the candidates
\begin{eqnarray}
\label{thetworoots}
\malpha_0^1 =  (1, -\omega^8)\,, \quad \malpha_0^2 = (1, -2 \omega^7)
\,.
\end{eqnarray}
We have already encountered algebras where a single new simple root
was not sufficient, but instead two new simple roots and a constraint
were necessary, in the case of $\widehat{E}_3$, and this is no
coincidence.  Let us calculate the mutual intersection $(\malpha_0^1,
\malpha_0^2)$.  In terms of explicit basis junctions as in
(\ref{junctioninbasis}), we have
\begin{eqnarray}
\malpha_0^1 = - \mz + \momega^q - \momega^8 \,, \quad \quad
\malpha_0^2 = - \mz + \momega^q - 2\momega^7 \,,
\end{eqnarray}
and then we calculate
\begin{eqnarray}
(\malpha_0^1, \malpha_0^2) &=& -1 + 2 \, (\momega^7, \momega^8) +
(\momega^q, \momega^q) \nn \\ 
&=& -1 - 6 +f (0,1) \\ %%%t = -7 + 7  \\ 
&=& 0 \,, \nn
\end{eqnarray}
and so we can take this
enhanced algebra to be the Kac-Moody algebra determined by the Dynkin
diagram in \figref{tworoots}(a), with an additional constraint on the
set of simple roots.   One
can compute that in $E_8$
\be
\omega^8 - 2\omega^7 = \alpha_1 + 2\alpha_2 +
3\alpha_3 + 2 \alpha_4 + \alpha_5 + 2 \alpha_8 - \alpha_7 \, .
\ee % 
It follows from this relation and (\ref{thetworoots})  
that
\begin{eqnarray}
\label{tworootconstraint}
\alpha_0^1 + \alpha_1 + 2 \alpha_2 + 3 \alpha_3 + 2 \alpha_4 +
\alpha_5 + 2 \alpha_8 = \alpha_0^2 + \alpha_7 \,.
\end{eqnarray}

It is possible to express this algebra as a further enhancement of an
affine semisimple algebra.  To see this, first remove the node
corresponding to $\alpha_6$ in the Dynkin diagram, giving us
\figref{tworoots}(b).  Without the two enhancing nodes
associated to $\alpha_0^1$ and $\alpha_0^2$, this is just the Dynkin
diagram of the semisimple algebra $E_6 \oplus A_1$.  The two enhancing
nodes then produce the affine enhancement,
$\widehat{E_6 \oplus A_1} = \widehat{E_6} \oplus \widehat{A_1}/\sim$,
just like the affinization $E_3 \ra \widehat{E}_3$. From this point
of view each
enhancing root satisfies $\alpha_0^i = -\theta_i + \delta$, where the
$\theta_i$ are
the highest roots of
$E_6$ and $A_1$ respectively and $\delta$ is the same in both cases.
Thus one can relate the two enhancing roots by
\begin{eqnarray}
\delta = \alpha_0^1 + \theta_{E_6} = \alpha_0^2 + \theta_{A_1} \,,
\end{eqnarray}
and using the $E_6$ and $A_1$ marks we find expressions for the
$\theta_i$ in terms of the $\alpha_i$ appearing in the Dynkin
diagram. This gives exactly (\ref{tworootconstraint}).

Thus Lie algebraically, the algebra we obtain by adding the enhancing
nodes to $E_8$ is the same as what we obtain starting with $E_6 \oplus
A_1$, affinizing, and then further enhancing by restoring the node
$\alpha_6$.  We can think of this as affinizing $E_8$ with respect to
its $E_6 \oplus A_1$ subalgebra.  In that sense, the new direction in
root space corresponds to the addition of a root $\delta$, which
satisfies $(\delta,\delta) =0$ and $(\delta,\alpha_i) = 0$ for all
roots in the $E_6 \oplus A_1$ subalgebra, but must satisfy
$(\delta,\alpha_6) = -1$ so that $(\alpha_0^i,\alpha_6)=0$.

%oliver  Here is just a comment. By Kac's book the Cartan matrix of
% the resulting algebra has zero determinant (gluing two affines by a
% new node) 
% but clearly the configuration does not admit a loop. The answer must be
% that the null eigenvector of the Cartan must be by our constraint the
% zero junction.
% If you agree and/or find this useful incorporate it.

This pattern of enhancements repeats itself for the other members of
the ${\bf E_N}$ series of configurations.  In general, the finite
$E_N$ algebra can be made to enhance to an indefinite algebra
which is an enhancement of $\widehat{E}_{N-2} \oplus
\widehat{A}_1/\sim$.
Alternately, this can be thought of as starting with $E_N$ and
affinizing with respect to the semisimple subalgebra $E_{N-2} \oplus
A_1$.  The appropriate values of $f(\mz)$ are the next-smallest after
those enhancing to ${\bf E_{N-1}^H}$.  In each case we find two Weyl
orbits satisfying (\ref{ll}) at $n=1$, giving two linearly dependent
enhancing simple roots.  We write the Cartan matrix by including both
new simple roots, and imposing a constraint on the set of roots.  Let
$c^i$ and $d^i$ be the extensions of the marks of $E_{N-2}$ and $A_1$
to $E_N$, defined as follows.  Let the $\alpha_i$ be the simple roots
of $E_N$. Then if $\alpha_i$ is a simple root of the $E_{N-2}$
subalgebra, $c^i$ is the mark of that $E_{N-2}$ root, otherwise
$c^i=0$, and likewise for $d^i$ and the $A_1$ subalgebra.  Then the
constraint on the simple roots for the enhanced algebra is
\begin{eqnarray}
\alpha_0^{E_{N-2}} + \sum_{i=1}^{N} \, c^i \, \alpha_i =
\alpha_0^{A_1} + \sum_{i=1}^N \, d^i \, \alpha_i \,,
\end{eqnarray}
where $\alpha_0^{E_{N-2}}$ and $\alpha_0^{A_1}$ are the two new simple
roots associated to affinizing the two different simple subalgebras.

\subsection{Rank 2 hyperbolic algebras} 

 There is a final interesting
family of enhancements we mention before ending this section.  ${\bf
E_1}$ can be enhanced to configurations realizing any simply-laced
rank 2 algebra with Cartan matrix
\begin{eqnarray}
A = \pmatrix{2 & -a
\cr -a & 2} \,,
\end{eqnarray}
for any positive integer $a$.  For ${\bf E_1}$, the quadratic form is
$f(p,q) = \fracs18 (p^2 + 2pq - 7q^2)$.  Junctions only arise for $p+q
= 0$ (mod 2) due to the charges of $\mB \mC \mC$.  Consider adding a
brane with charges $\mz = [\pm 2a - 1,1]$. 
%%%t This choice is necessary for interesting enhancement since
%%%t otherwise the new brane would not support the asymptotic charges
%%%t of ${\bf E_1}$. 
We obtain $f(\mz) = \fracs12 a^2 - 1$, and hence for $\mJ^2 = -2$ at
$n=1$ we find $\ll = \fracs12 a^2$. This is the length-squared of the
Weyl orbit of $su(2)$ with dominant weight given by Dynkin label $a$.
We are thus led to choose $\malpha_0 = (1, -a \omega)$, and
$(\malpha_0, \malpha) = -a$.

\begin{table}[t]
\begin{center}
\begin{tabular}{|c|c|c|c|c|c|} \hline
$\rule{0mm}{5mm}\gg$ & Config &Branes & $f(\mz)$ & $\genh$ &
Enh. branes \\
\hline \hline
 &&  & $-1/N$ & $A_{N+1}$ & $\mA{\bf H_N} $  \\
$A_N$ & ${\bf H_N}$ & $\mA^{N+1} \mC$ & $(N\hskip-2pt -\hskip-2pt
3)/(N\hskip-2pt
+\hskip-2pt 1)$ &
$D_{N+1}$ &
${\bf H_N}\mX_{\bf [3,1]}$ \\
 & &  & $(2N\hskip-2pt -\hskip-2pt 7)/(N\hskip-2pt +\hskip-2pt 1)$
 & $E_{N+1}$ & ${\bf H_N} \mX_{\bf [4,1]}$
\\
\hline
$D_N$ & ${\bf D_N}$& $\mA^N \mB \mC$ & $0$ & $D_{N+1}$ & $\mA{\bf D_N}$ \\
  &&   &  $(N-4)/4$ & $E_{N+1}$ & ${\bf D_N} \mC$ \\ \hline
$A_1$ &${\bf E_1}$& $\mB\mC\mC$ & $a^2/2 - 1$ & $\tilde{A_2}^{(a)}$ &
${\bf E_1}\mX_{\bf [2a-1,1]}$ \\ \hline
$\rule{0mm}{5mm}D_{4}$ & ${\bf D_{4}}$ & ${\bf A}^{4}{\bf BC}$ & $0$ &
$D_{5}$ & ${\bf D_4} \mX_{\bf [p,q]}$ \\ \hline
 & & & $0$ & $D_6$ & $\mA{\bf D_5}$\\
$D_5$ &${\bf D_5}$ & $\mA^5\mB\mC$ & $1/4$ & $E_6$& ${\bf D_5}\mC$  \\
 & $\cong {\bf E_5}$&& $1$ & $\widehat{D}_5$ & ${\bf D_5} \mX_{\bf [1,2]} $\\
&& & $9/4$ & $ E_4^H$ & ${\bf D_5} \mX_{\bf [1,3]}$ \\\hline
  &   && $1/3$ & $E_7$& $\mA{\bf E_6}$\\
$E_6$ &${\bf E_6}$& $\mA^5\mB\mC\mC$    & $1$ & $\widehat{E}_6$ & ${\bf
E_6}\mX_{\bf
[3,1]}$\\
     &   && $7/3$ & $ E_5^H$& ${\bf E_6} \mX_{\bf [4,1]}$\\ \hline
     &   && $1/2$ & $E_8$& $\mA{\bf E_7}$\\
$E_7$ &${\bf E_7}$  & $\mA^6\mB\mC\mC$ & $1$ & $\widehat{E}_7$&
${\bf E_7}\mX_{\bf [3,1]}$\\
   &     && $5/2$ & $ E_6^H$ & ${\bf E_7} \mX_{\bf [4,1]}$ \\ \hline
$E_8$ &${\bf E_8}$   & $\mA^7\mB\mC\mC$     & $1$ & $\widehat{E}_8 = E_9$
& $\mA{\bf E_8}$\\
  & && $3$ & $E_7^H$& ${\bf E_8} \mX_{\bf [4,1]}$\\ \hline
\end{tabular}
\end{center}
\caption{Finite algebras $\gg$ and enhanced
algebras $\genh$ obtained by adding a
single $\mz = [p,q]$ brane.  The Dynkin diagram of $\tilde{A}_2^{(a)}$
has two nodes  with $a$ lines joining them.
\label{enhancements}}
\end{table}

For $a=0$ this enhances to $A_1 \oplus A_1$, and for $a=1$ we find
$A_2$.  The value $a=2$ gives the affine algebra $\widehat{A_1}$, and
$a \geq 3$ gives the complete series of strictly hyperbolic simply
laced algebras of rank 2.  Imaginary roots appear for $a\ge 2$.  All
of these configurations only support asymptotic charges with $p+q = 0$
(mod 2).  All charges only become possible when an enhancing brane is
added with $[p,q]$ labels that do not satisfy this condition.

\bigskip

These cases only scratch the surface of the possible indefinite
algebras arising on 7-branes.  Ignoring for the moment that F-Theory
compactifications permit only 24 7-branes, and considering only the
algebraic properties of the monodromies, algebras of arbitrarily large
rank will appear in our construction. Most of these algebras have no
known relation to physics.  We do not, however, expect all
simply-laced algebras to arise.  For example, recall that there was no
way to enhance $D_4$ to $\widehat{D}_4$.  This affine algebra has the
simplest Dynkin diagram which contains four nodes each with a one line
connecting to a single, fifth node.  If this configuration appeared as
a subdiagram in any larger algebra we would expect to be able to
decouple branes and eventually reach $\widehat{D}_4$, so this must not
occur.  Thus any node in a Dynkin diagram arising on 7-branes must
branch to at most three other nodes.  In general the affine
$\widehat{A}_N$ and $\widehat{D}_N$ series do not appear (except for
$\widehat{A}_4 = \widehat{E}_4$ and $\widehat{D}_5 = \widehat{E}_5$),
so presumably they do not arise as subdiagrams either.  The complete
classification of indefinite algebras realized on 7-branes is still an
open problem.  The various enhancements obtained in the last three
sections are summarized in Table~\ref{enhancements}.

%%%%%%%%%%%%%%%%%%%%%%%%%%%%%%%%%%%%%%%%%%%%%%%%%%%%%%%%%%%%%%%%%%%

%\newpage
\newsection{$\sl2z$ Conjugacy Classes and Classification}
\label{classification}

In the previous sections we explored the appearance of various
infinite-dimensional algebras on configurations of 7-branes.  Which
algebra appears is largely determined by the total monodromy of the
brane configuration.  Since the physical situation is invariant under
global $\sl2z$ transformations, the Lie algebra is not associated with a
monodromy but rather with the conjugacy class of the monodromy. The 
monodromy is conjugated by this transformation,
and as a result we expect that the equivalence classes of 7-brane
configurations, and therefore the corresponding Lie algebras, are
related to the conjugacy classes of $\sl2z$.  It is this relation that
we explore in this section.

As we have commented, the equivalence class of the monodromy has the
dominant role in determining the algebra, but we are aware of two
other factors which must also be specified: the number of branes and
the integer $\mQ$ discussed in \secref{reviewsection}.  The number of
branes must be specified because there can exist sub-configurations of
branes with unit monodromy.  As we have proven, such configurations
must have some multiple of twelve branes, and specifying the total
number of branes removes the ambiguity they create.  The integer $\mQ$
is invariant under global $\sl2z$ and branch cut relocation, and
measures whether there are constraints on the possible asymptotic
charges.  Configurations with the same monodromy and number of branes
must still be inequivalent when $\mQ$ differs.  For almost all
configurations, $\mQ=1$, and all asymptotic charges can be realized.

Conjugacy classes of $\sl2z$ are conveniently organized by their
trace $t$.  Classes with $|t| < 2$ are known as elliptic, classes
with $|t| = 2$ parabolic and classes with $|t| > 2 $ hyperbolic.

In \cite{finite}, the finite Lie algebras arising on 7-branes were
matched to the corresponding $\sl2z$ conjugacy classes. 
%%%t
It is instructive to see how the rank of these algebras is fixed by
the monodromy. Consider for example, $t=0$, the representatives of the
two conjugacy classes at this trace are $S$ and $-S$. As follows from
the abelianization of $\sl2z$ (see \myref{listconj} and the discussion
below it), the configurations realizing the overall monodromy which is
conjugate to $S$ or $-S$ must contain \mbox{$3$ (mod 12)} and
\mbox{$9$ (mod 12)} 7-branes, respectively. In the simplest case of no
more than 12 branes this yields algebras of rank 1 ($A_1$) and 7
($E_7$), respectively. Similar arguments apply at other values of $t$.
%%%t 

There are two
classes for each elliptic $t$, and the six inequivalent elliptic
configurations were the three exceptional configurations ${\bf E_6}$,
${\bf E_7}$, ${\bf E_8}$ and the Argyres-Douglas ${\bf H_0}$, ${\bf
H_1}$ and ${\bf H_2}$, all collapsible configurations leading to
Kodaira singularities. There are an infinite number of classes at
both $t=2$ and $t=-2$, and some of these correspond to configurations of
mutually local branes ${\bf A_N}$ ($t=2$) which are collapsible for
all $N$, and ${\bf D_N}$ configurations ($t=-2$), which exist for all
non-negative $N$ but are collapsible only for $N \geq 4$.  These
series fill an infinite number of parabolic conjugacy classes, but an
infinite number still remain.  Finally, the ${\bf E_N}$ configurations
could be extended to $0 \leq N \leq 5$, and the ${\bf H_N}$ to $N \geq
3$.  The cases ${\bf E_5} \cong {\bf D_5}$ and ${\bf H_3} \cong {\bf
D_3}$ correspond to parabolic classes, while the rest are
associated to hyperbolic classes of negative trace.  At
$t=-6$ we had the conjugacy class for ${\bf E_1}$
and ${\bf \tilde{E}_1}$, these configurations are inequivalent since  
$\mQ({\bf E_1}) = 2$ and $\mQ({\bf \tilde{E}_1}) = 1$.  These results 
are summarized in table \ref{maintable}.

Let us now explore how the infinite-dimensional algebras fit into this
scheme.  They can occupy the remaining parabolic and hyperbolic
conjugacy classes.  The number $H(t)$ of hyperbolic conjugacy classes
at each value of $t$ cannot be determined by elementary methods as was
done for the elliptic and parabolic cases in \cite{finite}.  There one
could relate the fixed point of the elements of an elliptic or
parabolic class to a fixed point in the fundamental domain ${\cal F}$,
but since the fixed points of hyperbolic classes are irrational real
numbers, they cannot be mapped to ${\cal F}$.  Instead, more
sophisticated machinery is necessary to determine $H(t)$ for $|t| >2$.

One determines $H(t)$ by using the isomorphism between the conjugacy
classes
of $\sl2z$ matrices of trace $t$ and equivalence classes of binary
quadratic forms of discriminant $t^2-4$.  This isomorphism associates
the charge quadratic form with a given monodromy, as discussed in
\cite{finite}.  The number $H(t)$ of conjugacy classes in $\sl2z$ with
trace $t$ is
\begin{equation}
H(t)=h_+(t^{2}-4),
\end{equation}
where $h_{+}(\widehat{d})$ denotes the number of equivalence classes
of binary quadratic forms with discriminant $\widehat{d}$. Moreover,
the equivalence classes of binary quadratic forms of discriminant
$\widehat{d}$ are in one-to-one correspondence with the (strict)
equivalence classes of ideals in a quadratic field of discriminant
$\widehat{d}$.  This relation allows us to determine $H(t)$ for
arbitrary $t$.  This process is somewhat technical, and is explained
in the appendix.  There table \ref{numberofclasses} lists the number
of conjugacy classes and gives representatives for $-15\leq t \leq
15$.

In section \ref{affinize} we found configurations ${\bf
\widehat{E}_N}$ realizing affine versions of the exceptional algebras.
We noted there that all $\sl2z$ matrices $K$ with a unit eigenvalue
have $t = {\Tr} K = 2$, including the monodromy matrices $K({\bf
\widehat{E}_N})$ of the affine configurations:
\begin{eqnarray}
{\Tr} K({\bf\widehat{E}_N)} = 2 \,.
\end{eqnarray}
Thus all the affine algebras are associated to parabolic conjugacy
classes.  Representatives of the $t=2$ classes are $\left({1\,
-\!n}\atop{\!\!0\;\;\; 1}\right)$ with $n \in \bbbz$.  From
(\ref{Kaffine}), we see that ${\bf \widehat{E}_N}$ configurations
correspond to classes with $n = N-9$.

At $N=1$, there are the two inequivalent configurations ${\bf
\widehat{E}_1}$ and ${\bf \widehat{\tilde{E}}_1}$ which are the
affinizations of the corresponding inequivalent finite cases as explored 
in \cite{finite}.  They
share the same conjugacy class, and are distinguished by $\mQ$.  For
$N=0$ we have the ${\bf \widehat{\tilde{E}}_0}$ configuration.  Thus
${\bf \widehat{E}_N}$ is defined for all $N \geq 0$.  For $N>9$ the
monodromy of this series coincides with that of ${\bf A_{N-10}}$, as
noted at the end of section \ref{e9hat}; the different series differ
by a factor of ${\bf \widehat{E}_9}$.

The series ${\bf E_N}$ realizes infinite algebras for $N \geq 9$.  For
$N=9$ this is just the affine configuration ${\bf E_9} \cong {\bf
\widehat{E}_8}$, while for $N \geq 10$ we have a series of infinite-
dimensional exceptional algebras.  From Table \ref{series} we have
\begin{eqnarray}
{\Tr} K({\bf E_N}) = N - 7 \,,
\end{eqnarray}
we see that for $N \geq 10$ the associated conjugacy classes
are hyperbolic with positive trace.

In section \ref{indefinite}, we found the hyperbolic exceptional
configurations ${\bf E^H_N}$.  Using (\ref{hypK}) one can see 
\begin{eqnarray}
{\Tr} K({\bf E_N^H}) = 11 - N \,.
\end{eqnarray}
At $t=3$, we find $H(3) = 1$ and thus a single conjugacy class for
${\bf E_{10}} \cong {\bf E^H_8}$.  The other elements in the ${\bf
E_N}$ and ${\bf E_N^H}$ series are distinct, and correspondingly $H(t)
> 1$ for $3<t<10$ where both are defined.  These configurations are
displayed up to $t=7$ in Table \ref{maintable}.

Section \ref{indefinite} discussed a few other configurations
realizing infinite-dimensional algebras which lie outside the scope of
Table \ref{maintable}.  The enhancement of ${\bf E_8}$ to a
configuration requiring two new simple roots, with Dynkin diagram
given in figure \ref{tworoots}, appears at $t=8$.  Since ${\bf
E_{15}}$ and ${\bf E_3^H}$ are already expected to be present as well,
we anticipate that there are more than two conjugacy classes at $t=8$,
and indeed $H(8) = 4$.  The corresponding enhancements of ${\bf E_7}$,
${\bf E_6}$ and ${\bf E_5}$ occur at $t=10,12$ and $14$, respectively.

We also considered the enhancement of ${\bf E_1}$ to the
general series of rank 2 simply laced algebras.  These configurations
are associated to the algebras $\tilde{A}_2^{(a)}$ with Dynkin diagram
$A = \left({\;2\;-\!a}\atop{\!\!\!-\!a\;\;\,2}\right)$, and have trace
\begin{eqnarray}
{\Tr} K = 4a^2 - 14 \,.
\end{eqnarray}
The $a=0$ and $a=1$ configurations with algebras $A_1 \oplus A_1$ and
$A_2$ have $t = -14$ and $t=-10$, joining the other exotic finite
algebras in the hyperbolic classes of negative trace.  The case with
$a=2$ is just ${\bf \widehat{E}_1}$ and has $t=2$ as expected, and
thereafter the algebras are strictly hyperbolic and the configurations
appear at large positive values of $t$, $a=3$ being at $t=22$.  Even
more exotic configurations, which we have not explored here,
presumably populate the higher-$t$ classes.

We found configurations realizing the parabolic classes with
$\left({1\,-\!N}\atop{\!\!0\;\;\;1}\right)$ for $N \geq -9$ (${\bf
\widehat{E}_{N+9}}$ and ${\bf A_{N-1}}$) and
$\left({\!-\!1\,N}\atop{0\,-\!1}\right)$ for $N \geq -4$ (${\bf
D_{N+4}}$).  It is natural to wonder about the remaining classes.  One
moves down the ${\bf \widehat{E}_N}$ and ${\bf D_N}$ series by
removing $\mA$-branes, but at the ends of the series there are no more
such branes to remove.  However, we can add to the ${\bf
\widehat{E}_0}$ and ${\bf D_0}$ configurations a set of branes with
total monodromy inverse to that of a single $\mA$-brane, namely the
configuration with trivial monodromy ${\bf \widehat{E}_9}$ with a
single $\mA$-brane removed, in other words ${\bf \widehat{E}_8} =
\mA^7 \mB \mC \mB \mC$.  One can then remove $\mA$ branes to continue
into the other parabolic classes, adding ${\bf \widehat{E}_8}$ factors
when necessary.  In this fashion we exhaust the parabolic classes.  We
do not attempt to classify the algebras on these configurations.

As an example consider the ${\bf D_{N}}$ series. ${\bf D_{0}=BC}$ and
has no ${\bf A}$-branes which can be removed. As far as the monodromy
is concerned ${\bf D_{0}}$ is equivalent to ${\bf
\widehat{E}_{9}\,D_{0}}= \mA^{8}\mB\mC\mB\mC\mB\mC$, which has ${\bf
A}$-branes which can be removed to obtain, for example, ${\bf D_{-1}}
\equiv \mA^{7}\mB\mC\mB\mC\mB\mC$. Conjugacy classes
$\left({\!-\!1\,N}\atop{0\,-\!1}\right)$ for $N < -4$ are thus
realized by 7-brane configuration made of ${\bf A}$-branes and an odd
number of ${\bf BC}$ pairs.

In general, an ${\bf \widehat{E}_9}$ configuration can be added to 
any any configuration $\mG$ without changing the monodromy:
\begin{eqnarray}
K(\mG) = K(\mG {\bf \widehat{E}_9}) \,,
\end{eqnarray}
and consequently the entire classification possesses a certain 
${\bf \widehat{E}_9}$-periodicity.  To any classified configuration
$\mG$ with $n$ branes, we can add $k$ factors of ${\bf \widehat{E}_9}$
to obtain a new configuration with $n+12k$ branes and algebra
\begin{eqnarray}
\gg(\mG \, {\bf \widehat{E}_9 \, \widehat{E}_9} \, \cdots \, {\bf
\widehat{E}_9}) \supseteq \gg(\mG) \oplus \widehat{E}_9 \oplus \cdots
\oplus \widehat{E}_9 \,.
\end{eqnarray}
In general we have made no attempt to classify these algebras.  A few
simple cases can be understood; for example $\mA {\bf \widehat{E}_9} =
{\bf \widehat{E}_{10}}$.  The progression of the ${\bf E_N^H}$
series in Table \ref{maintable} suggests that ${\bf E_{10}^H} \cong
{\bf H_0 \widehat{E}_9}$, and so on.

The algebraic facts that $E_9 = \widehat{E}_8$ and $E_{10} = E^H_8$
are reflected in equivalence of the corresponding the brane
configurations.  In Table \ref{e8enhance} we depict how beginning with
${\bf E_8}$, we can add an $\mA$-brane to proceed along the ${\bf
E_N}$ series, or an $\mX_{\bf [3,1]}$-brane to move toward loop and
then hyperbolic algebras.  The table is reflection-symmetric along the
diagonal as a result of these equivalences.  Because ${\bf E_9^H \cong
\widehat{E}_{10}}$, this suggests that we can define the algebra
$E_9^H$ to be equal to $\widehat{E}_{10}$, which is just the loop
algebra of $E_{10}$. Notice the central position occupied by the
double-loop configuration ${\bf \widehat{E}_9}$.

In Table \ref{maintable} we summarize this discussion, presenting the
algebras arising on a collection of 7-branes with total monodromy of
trace $|t|\leq 7$.

\begin{table}
\begin{center}
\begin{tabular}{|c|c|c|c|l|c|}
\hline
$\rule{0mm}{5mm}$ $K$-type & ${\Tr}K$ & det $A(\gg)$ &
   Brane configuration $\mG$ & $\hspace{.35in}K$ & $H(t)$ \\ \hline
\hline
&$\rule{0mm}{7mm}$$-7$ &9& $ {\bf \tilde{E}_0}\,,{\bf H_8} $
&$\pmatrix{-7 & 1\cr -1& 0}^{\pm1}   $  & 2
\\ \cline{2-6}
&$\rule{0mm}{7mm}$$-6$ &(2,8),8& $ ({\bf E_1}, {\bf \tilde{E}_1})
                                        \,, {\bf H_7}  $
&$\pmatrix{-6 & 1\cr -1& 0}^{\pm1}  $  & 2
\\ \cline{2-6}
$hyp.$
&$\rule{0mm}{7mm}$$-5$ & 7 & $ {\bf E_2}\,, {\bf H_6}  $
&$\pmatrix{-5 & 1\cr -1& 0}^{\pm1}  $  & 2
\\ \cline{2-6}
&$\rule{0mm}{7mm}$$-4$ & 6 & $ {\bf E_3}\,, {\bf H_5}  $
&$\pmatrix{-4 & 1\cr -1& 0}^{\pm1}  $  & 2
\\ \cline{2-6}
&$\rule{0mm}{7mm}$$-3$ & 5 & $ {\bf E_4}  = {\bf H_4}  $
&$\pmatrix{-3 & 1\cr -1& 0}  $  & 1
\\ \hline
$par.$
&$\rule{0mm}{8mm}$$-2$ &4  &
$\begin{array}{c} {\bf D_{N+4 \,\geq 0}} \\
 ({\bf E_5}\!=\!{\bf D_5}\,,{\bf H_3}\!=\!{\bf D_3})\end{array}$
&$ \pmatrix{-1 & N\cr  0&-1} $  & $\infty$
\\ \hline
&$\rule{0mm}{5mm}$$-1$& 3&${\bf E_6}\,,{\bf H_2}$
&$  ~-\!(ST)^{\mp1}$&2
\\ \cline{2-6}
$ell.$
&$\rule{0mm}{5mm}$ 0& 2&${\bf E_7}\,,{\bf H_1}$
&$  ~~~~~~S^{\pm1}    $&2
\\ \cline{2-6}
&$\rule{0mm}{5mm}$ 1& 1&${\bf E_8}\,,{\bf H_0}$
&$ ~~~~(ST)^{\pm1}   $&2
\\ \hline
$par.$
&$\rule{0mm}{8mm}$ 2 & $\begin{array}{c} N\\ 0\end{array}$
&$\begin{array}{c} {\bf A_{N-1\,\geq 0}}\\
           {\bf \widehat{E}_{N+9\geq 0}},{\bf\widehat{\tilde{E}}_{1}},
           ({\bf E_9}={\bf \widehat{E}_8}) \end{array}$
& $\pmatrix{ 1 &-N\cr  0& 1}$ & $\infty$
\\ \hline
&$\rule{0mm}{7mm}$ 3 &  $-1 $& $ {\bf E_{10}}  = {\bf E_8^H}$
& $\pmatrix{ 0 & 1\cr -1&3 }$ & 1
\\ \cline{2-6}
&$\rule{0mm}{7mm}$ 4 & $ -2 $& $ {\bf E_{11}}\,, {\bf E_7^H}$
& $\pmatrix{ 0 & 1\cr -1&4 }^{\pm1}$ & 2
\\ \cline{2-6}
$hyp.$
&$\rule{0mm}{7mm}$ 5 & $ -3 $& $ {\bf E_{12}}\,, {\bf E_6^H}$
& $\pmatrix{ 0 & 1\cr -1&5 }^{\pm1}$ & 2
\\ \cline{2-6}
&$\rule{0mm}{7mm}$ 6 & $ -4 $& $ {\bf E_{13}}\,, {\bf E_5^H}$
& $\pmatrix{ 0 & 1\cr -1&6 }^{\pm1}$ & 2
\\ \cline{2-6}
&$\rule{0mm}{7mm}$ 7 & $ -5 $& $ {\bf E_{14}}\,, {\bf E_4^H}$
& $\pmatrix{ 0 & 1\cr -1&7 }^{\pm1}$ & 2 \\ \hline
\end{tabular}
\end{center}
\caption{$\sl2z$ conjugacy classes and algebras realized on 7-branes
with overall monodromy $K$ (up to conjugation). The upper and lower
exponents of the matrices correspond to the first and the second brane
configuration, respectively. The brane configurations $\mG$, whose
notations suggest the algebra $\gg$, are defined in the text. The
determinant of $A(\gg)$ is given, except when no algebra is realized,
as with ${\bf H_0,\tilde{E}_0}$ and ${\bf D_0}$.  \label{maintable}}
\end{table}

\begin{table}
\begin{center}
\begin{tabular}{||c|c|c||}
\hline
$\rule{0mm}{7mm}
{\bf E_8^H} = {\bf E_8}\mX^{\bf 2}$
&  ${\bf E_9^H} = \mA{\bf E_8} \mX^{\bf 2}$
& ${\bf E_{10}^{H}} = \mA^{\bf 2}{\bf E_8}\mX^{\bf 2}$  \\
\hline
$\rule{0mm}{7mm}
{\bf \widehat E}_{\bf 8} \,=\, {\bf E_8 X}$
& ${\bf \widehat{E}_9} = \mA{\bf E_8}\mX $&
 ${\bf \widehat{E}_{10} } = \mA^{\bf 2}{\bf E_8} \mX $ \\
\hline
$\rule{0mm}{7mm}{\bf E_8} = \mA^{\bf 7}\mB\mC^{\bf 2}$
& ${\bf E_9 }  = \mA {\bf E_8}$ &
${\bf E_{10}} = \mA^{\bf 2}{\bf E_8} $
\\ \hline
\end{tabular}
\end{center}
\caption{Enhancements of the ${\bf E_8}$ brane configuration. In the
horizontal direction we add $\mA$ branes and in the vertical direction
we add $\mX = [3,1]$ branes. The brane configurations produce the
algebras suggested by their names.  This square is reflection
symmetric with respect to the diagonal running from the lower left
corner to the upper right corner.
\label{e8enhance}}
\end{table}

\vfill
\break
\goodbreak
\newsection{Compactifications of type IIB superstrings}
\label{compactifications}

Throughout this paper we have investigated the algebraic properties of
the monodromies associated with 7-branes, without considering the
physics of the background the 7-branes are realized in.  The
well-known string theory background that inspired this work is that of
IIB strings compactified on $S^2$ in the presence of 24 mutually
non-local 7-branes.  This can also be viewed as a compactification of
F-Theory on an elliptic K3, where the fibration over $\bbbp^1$
degenerates at 24 points.  In this section we address the question of
which algebras actually appear in such a background. Since the finite
algebras have already been studied in some detail
\cite{mgthbz,lerche1,lerche2}, our focus here is on the infinite
algebras. We find configurations of branes realizing many of the
affine and indefinite algebras discussed in this paper, existing in
certain regions of the moduli space of compactifications.  As
emphasized, infinite-dimensional algebras are not associated to
singularities, since these brane configurations cannot coalesce
(unless the whole K3 degenerates). Thus, the associated theories are
always spontaneously broken, and while the massive spectrum does fall
into representations of the algebra, states belonging to any
particular representation will not in general have the same mass.  An
additional subtlety is that owing to the global properties of the
base, loop junctions around the entire brane configuration can be
contracted to a point and are actually trivial.

Another interesting manifold with an elliptic fibration structure is
the ninth del Pezzo surface ${\cal B}_9$, also called $\half$K3 since
it is a fibration over $\bbbp^1$ degenerating at 12 points.  This
manifold is not Calabi-Yau and so is not a candidate for F-Theory
compactification, but arises as a 4-cycle that can collapse to zero
size inside certain Calabi-Yau threefolds.  These collapsing 4-cycles
lead to interesting theories with tensionless strings in 6D and 5D.
D2-branes wrapped on the holomorphic curves of this manifold lead to
electrically charged BPS particles in the $\cal{N}$=2, $D=4$ theory in
the transverse space of a IIA compactification
\cite{klemm,douglas,warner}. The spectra of BPS states can be
reproduced by looking at BPS junctions of asymptotic charge $(p,0)$ on
the ${\bf \widehat{E}_{9}}$ configuration. The degree of a holomorphic
curve in ${\cal{B}}_9$ is then equal to the the $p$-charge of the
corresponding junction. The 2-cycles in ${\cal B}_9$ and their
intersection matrix can be determined using the techniques of this
paper, even though they do not correspond to junctions of $(p,q)$
strings.  The ``brane'' configuration on this surface will be seen to
be ${\bf \widehat{E}_9}$, although the global properties of the base
will again render certain junctions trivial.

\subsection{Junctions and the homology lattice of ${\cal B}_9$}  

${\cal B}_9$ is a complex
manifold which can be expressed as an elliptic fibration over the base
$B \cong \bbbp^{1}$ \cite {kehagias, witten/donagi, vafawarner,
ganor}.  The Euler characteristic of ${\cal B}_9$ is twelve and hence
it has twelve degenerate fibers. The positions of the degenerate
fibers on the base $B$ are the positions of the twelve
``7-branes''. ${\cal B}_9$ can also be obtained by blowing up nine
generic points on $\bbbp^{2}$. The homology lattice of ${\cal B}_9$
has as a basis the cycles $l,e_{1},\ldots,e_{9}$, where $l$ is basic
homology class of $\bbbp^{2}$ and $-e_{i}$ are the classes obtained by
blowing up the nine points. The intersection numbers of these classes
are given by

\begin{equation}
l^{2}=1,\hspace{.3in} l\cdot e_{i}=0,\hspace{.3in} 
e_{i}\cdot e_{j}=-\delta_{ij},\hspace{.3in}
\forall\ i,j\in \{1,\ldots,9\}.
\end{equation}
The homology lattice is isomorphic to the lattice $\Gamma^{9,1}=
-A(E_{8})\oplus \Gamma^{1,1}$, where $\Gamma^{1,1}
=\left({\!\!1\;\;0}\atop{0\,-\!1}\right)$. This can be seen by
defining a new basis
\begin{eqnarray}
\alpha_{i}&=& e_{i}-e_{i+1},\hspace{.8in} 1\leq i \leq 7, \nn \\
\alpha_{8}&=& l-e_{1}-e_{2}-e_{3}, \\
B &=& e_{9}, \nn \\
F &=&3l-\sum_{i=1}^{9}e_{i}. \nn
\end{eqnarray}
The classes $B$ and $F$ represent the fiber and the base of the
elliptic fibration. In this basis $\alpha_{i},i=1,\ldots,8$ have the
intersection matrix $-A(E_{8})$. $B+F$ and $B$ generate the basis of
the lattice $\Gamma^{1,1}$ and have the intersection matrix
$\left({\!1\;\;0}\atop{0\,-\!1}\right)$. It is not possible to get
the hyperbolic lattice $\left({ 0\;\;1}\atop{1\,\,\,\,\!0}\right)$
by a change of basis which preserves the fact that the basis should be
integral homology classes,  and therefore it follows that the ten
dimensional homology lattice of ${\cal B}_9$ is not the root lattice
of $E_{10}$.

Since the base of the elliptic fibration of ${\cal B}_9$ is compact,
the total monodromy of the ``7-brane'' configuration must be unity,
since the branch cuts have nowhere to go.  Since there are twelve
branes, the configuration in question must be ${\bf \widehat{E}_9}$.
The full junction lattice on ${\bf \widehat{E}_9}$ is
twelve-dimensional.  The condition of vanishing asymptotic charge is
equivalent, on this compact surface, to the condition of zero
boundaries on the curves.  The remaining ten-dimensional lattice is
further reduced by the fact that since the 7-branes are on a sphere,
any loop around the entire configuration can be shrunk to a point.
This means that both imaginary root junctions must be set equal to the
zero junction:
\begin{eqnarray}
\mdelta_1 = \mdelta_2 =0 \,.
\end{eqnarray}
The remaining eight-dimensional lattice is just $-A(E_8) \cong \subset
\Gamma^{9,1}$, corresponding to the $A(E_8)$ part of the ${\cal B}_9$
junction lattice.  Since the junctions always have one dimension in
each the base and the fiber, the cycles corresponding to wrapping
completely either the base or the fiber are invisible in the junction
picture.

\subsection{Junctions and the homology lattice of K3}

The second homology group of K3 is a 22-dimensional lattice of
signature $(19,3)$: $\!H^{2}(\mbox{K3},\!\bbbz)\!\!=\Gamma^{19,3}$.  In
a particular basis the intersection matrix is given by
\begin{equation}
[-A(E_{8})]\oplus [-A(E_{8})] \oplus H\oplus H\oplus H \,,
\end{equation}
where $A(E_{8})$ is the Cartan matrix of $E_{8}$
and $H=\left({0 \,1}\atop {1\,0}\right)$. Out of the three $H$-factors,
one is the intersection matrix of $[F]$ and $[B]+[F]$, where $[B]$ is
the homology class of the base and $[F]$ is the homology class of the
fiber.

Even though we have twenty four branes on the two-sphere associated to
the K3, a junction now cannot have asymptotic charge and therefore the
junction lattice is at most twenty two dimensional. In addition, since
the total monodromy around the entire configuration is trivial there
are two linearly independent junctions encircling all the
branes. These junctions can be shrunk to a point on the other side of
the base $B \cong \bbbp^{1}$ and are therefore trivial. As a
consequence, the lattice of junctions is twenty dimensional.
%%% The full junction lattice realized on 7-branes is twenty-four
%%% dimensional. The sub-lattice of junctions with zero asymptotic
%%% charge has dimension twenty-two. Since the total monodromy around
%%% the entire configuration is trivial, we can have two linearly
%%% independent junctions encircling the twenty-four 7-branes. But
%%% again since these junctions can be shrunk to a point on the other
%%% side of the base $B\cong \bbbp^{1}$, these loops are equivalent to
%%% the trivial junction and the corresponding 2-cycles are zero in
%%% homology. Thus the sub-lattice of junctions of zero asymptotic
%%% charge is actually 20-dimensional and 
It corresponds to the lattice
$\Gamma^{18,2}=[-A(E_{8})] \oplus [-A(E_{8})] \oplus H\oplus H$, the
sub-lattice of $H^2(\mbox{K3},\bbbz)$ generated by the classes other
than $[B]$ and $[F]$. \figref{K3} shows a basis of this sub-lattice,
with two ${\bf \widehat{E}_9}$ brane configurations, similar to a
basis used in \cite{aspinwall}. (We explain in the next section how
${\bf \widehat{E}_9}{\bf \widehat{E}_9}$ arises on the full K3.) The
part of the basis giving one of the $H$'s consists of $\delta_{3,1}$
and $\delta_{3,1}+AA$ where $\delta_{3,1}$ is the $(3,1)$-loop and
$AA$ stands for the string connecting the {\bf A}-branes. Similarly
$\delta_{1,0}$ and $\delta_{1,0}+XX$ generate the other $H$ and the
root junctions of the two $E_8$'s give $[-A(E_{8})] \oplus
[-A(E_{8})]$.

\onefigure{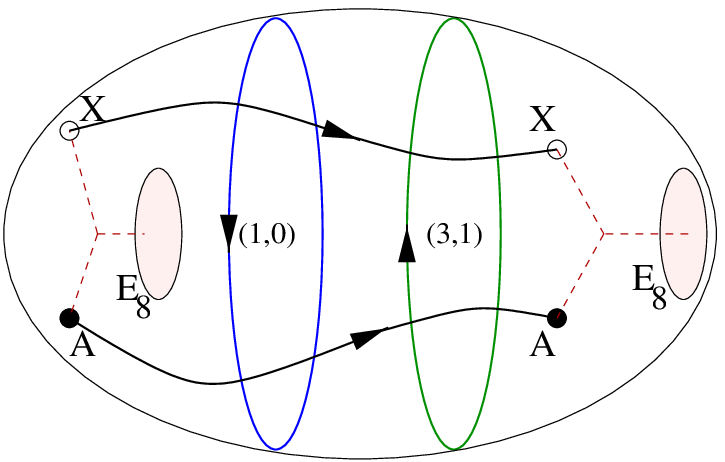}{The 24 7-branes of the full K3 are grouped to uncover
the $\Gamma^{18,2}=[-A(E_{8})] \oplus [-A(E_{8})] \oplus H\oplus H $
part of the homology lattice.}

\comment{ %%%%%%%%%%%%%%%%%%%%%%%%%%%%%%%%%%%%%%%%%%%%%%%%%%%%
{\bf Signature \mbox{(18,2)} from junction lattice.} Let us verify
that the signature of the lattice is $(18,2)$.  Consider the brane
configuration ${\bf (AC)^{12}}$ which is known to exist on K3.  An
explicit basis of the 20-dimensional lattice of nontrivial junctions
with no asymptotic charge is given by
\be
{\bf \alpha}_{i}={\bf a}_{i}-{\bf a}_{i+1},\,\,
{\bf \beta}_{i}={\bf c}_{i}-{\bf c}_{i+1},\,\,i=1,\ldots,10.
\ee

The various intersection numbers are 
\be
{\bf \alpha}_{i}\cdot{\bf \alpha_{j}}={\bf \beta}_{i}\cdot{\bf
\beta}_{j}= -A_{ij}(A_{10}),\,\,
{\bf \alpha}_{i}\cdot{\bf \beta}_{j}=\delta_{i,j}-\delta_{i-1,j}
\equiv B_{ij}.
\ee
The intersection matrix is then
\be
A(K3)=\pmatrix{-A(A_{10}) & B\cr B & -A(A_{10})}
\ee
This matrix has signature $(18,2)$ and thus corresponds to the 
intersection matrix of a basis of second homology not containing 
the classes corresponding to the base and the fiber.
}%%%%%%%%%%%%%%%%%%%%%%%%%%%%%  comment  %%%%%%%%%%%%%%%%%%%%%%%%%%%

\subsection{Brane configurations on the full K3}

Let us now examine configurations realizing infinite-dimensional
algebras on K3.  We can start with eight ${\bf H_1}$ singularities:
\begin{equation}
\label{fullmonty}
{\bf K3} = ({\bf AAC})({\bf AAC})({\bf AAC})({\bf AAC})\,\,\,\,
({\bf AAC})({\bf AAC})({\bf
AAC})({\bf AAC})\,.
\end{equation}
The total monodromy of this configuration is unity, as it should be.
Since the branch cuts all go downwards to a single point on the two
sphere and there are no more branes around, branes can move
cyclically. We now recall that ${\bf CAA = AAB}$. This allows one to
write \cite{mgthbz} 
\be
\label{mbrane}
{\bf AAC \, AAC=AAAA\, BC = D_4}\,.  \ee Thus the
full configuration is equivalent to \be {\bf K3} \cong {\bf
(A^4\, BC) (A^4\, BC)\, (A^4\, BC)\, (A^4\, BC) = D_4D_4D_4D_4 }\,,
\ee the familiar configuration giving rise to an $so(8)^4$ algebra
\cite{sen}.  The algebra on this configuration is even larger when we
take into account junctions that stretch between ${\bf D_4}$ factors.
We can now combine them in pairs and find \be
\label{enineenine}
{\bf K3} \cong (\mA^8\, \mB\mC\, \mB\mC)\, (\mA^8\, \mB\mC\,
\mB\mC) = {\bf \widehat E_9 \, \widehat E_9 } \,.  
\ee 
Let us, however impose global constraints.  Each ${\bf \widehat{E}_9}$
has a pair of associated $\mdelta$ junctions, one each of $(1,0)$ and
$(3,1)$ charges: $\mdelta^i_{\bf A}, \mdelta^i_{\bf X}$.  In fact
$\mdelta^1_{\bf A} + \mdelta^2_{\bf A}$ and $\mdelta^1_{\bf X} +
\mdelta^2_{\bf X}$ are the two junctions that wind around the entire
configuration, and hence are equivalent to zero.  We must impose a
condition on the basis strings,
\begin{eqnarray}
\label{deltaequiv}
\mdelta_{\bf A}^1 = -\mdelta_{\bf A}^2 \,, \quad \quad 
\mdelta_{\bf X}^1 = - \mdelta_{\bf X}^2
\,.
\end{eqnarray}
Thus instead of four nontrivial imaginary roots on the total algebra,
we have just two, which we can think of as the loops ``caught'' in
between the two brane configurations, as in \figref{K3}.  
%%%t
Therefore, an arbitrary junction will be characterized by two levels
only. The subalgebra readily identified here is therefore the quotient
$(\widehat{E}_9\oplus\widehat{E}_9)/(K_A^1+K_A^2,K_X^1+K_X^2)$, where 
$(K_A^1,K_X^1)$ and $(K_A^2,K_X^2)$ are the central elements of the
first and second $\widehat{E}_9$ respectively. Note that a weight
vector of this algebra is defined by $18 (=10\times 2-2)$ integers and
thus it cannot capture the full set of invariant charges (20 of them)
of the associated junction. 

Alternatively, we can associate one imaginary root in \figref{K3} to
one of the $E_8$'s and the other root to the other $E_8$ obtaining an
equivalent description in terms of an $E_9\oplus E_9$ subalgebra
without a quotient. In this case the failure to describe the full
junction lattice can be ascribed to the existence of two nontrivial
junctions that are $E_9\oplus E_9$ singlets. Indeed, consider the side
with brane configuration ${\bf A\widehat{E}_8}={\bf
AE_8X_{[3,1]}}$. The relevant singlet here is tadpole like, with the
loop surrounding the ${\bf \widehat{E}_8}$ (carrying one $E_9$) and
the leg a $(1,0)$-string ending on the ${\bf A}$-brane (recall the
discussion around \myref{jqqaffine}). Exactly analogous
considerations apply to the other side.

%%%t We can associate each imaginary root with one set of branes, thus
%%%t realizing an $E_9 \oplus E_9$.  This is not the whole story
%%%t however, because there are still junctions that can stretch
%%%t between one configuration and the other.

{\bf The $\bf E_{10}\oplus E_{10}$ configuration.} It is natural to
rearrange the branes so that it is the junctions stretching in between
the two configurations which are removed by the global constraints.
%%%t In fact we cannot do this completely, but we can come close.  
Rewrite the configuration ${\bf K3}$ in (\ref{enineenine}) as
\begin{eqnarray}
{\bf K3} &\cong& (\mA^8\, \mB\mC\mC\, \mX_{\bf [3,1]})\,\, (\mA^8\,
\mB\mC\mC \,\mX_{\bf [3,1]}) = (\mA {\bf E_8} \mX_{\bf [3,1]})\, ( \mA
{\bf E_8} \mX_{\bf [3,1]}) \,,
\end{eqnarray}
where each $( \mA {\bf E_8} \mX_{\bf [3,1]})$ has unit monodromy and
so all its branch cuts meet at a point.  Since  
${\bf AE_{8}X_{[3,1]}}={\bf X_{[3,1]}AE_{8}}$  we can  
move one $\mX_{\bf [3,1]}$ brane around the other configuration 
entirely, and regroup the branes:
\begin{eqnarray}
\label{eteneten}
{\bf K3} &\cong& (\mA {\bf E_8} \mA) \, ({\bf E_8} \mX_{\bf [3,1]}
\mX_{\bf [3,1]})
= {\bf E_{10} \,\, E_{10}' }\,. 
\end{eqnarray}
The brane configurations ${\bf AE_{8}A}$ and ${\bf
E_{8}X_{[3,1]}X_{[3,1]}}$ are both equivalent to ${\bf
E_{10}}$. Therefore both configurations realize the algebra
$E_{10}$. They have inverse monodromies, and there is a branch cut
stretching between them.  In $\sl2z$ a matrix and its inverse have the
same trace; since there is a single conjugacy class at trace equal
$(+3)$, the second configuration in fact must be conjugate (and
equivalent) to the first.

%%%t{\bf Singlets of $E_{10}$ and junction localization.} 
An interesting phenomenon of junction localization occurs for the
configuration \myref{eteneten}. Since $E_{10}$ has no conjugacy
classes a singlet can be realized on ${\bf E_{10}}$ with any
asymptotic charge. A singlet of $E_{10}\oplus E_{10}$ will then be a
loop around the first ${\bf E_{10}}$ with some asymptotic charge
$(p,q)$ and another loop around the second ${\bf E_{10}}$
configuration with asymptotic charge $(-p,-q)$, as shown in
\figref{e10}(a). From the way singlets are constructed \cite{finite}
it is clear that this singlet can be realized as shown in
\figref{e10}(b) and is thus zero in homology because it can be
contracted to a point on the other side of the sphere. Consider a
generic junction $\mJ$ stretching between the two ${\bf E_{10}}$
configurations taking an asymptotic charge $(p,q)$ from one ${\bf
E_{10}}$ to the other ${\bf E_{10}'}$. By adding a suitable singlet we
can eliminate the string going between the configurations. Thus any
junction $\mJ$ is equivalent to the sum of two junctions $\mJ_{1}$ and
$\mJ_{2}$ localized on the first and the second configuration
respectively.  
\onefigure{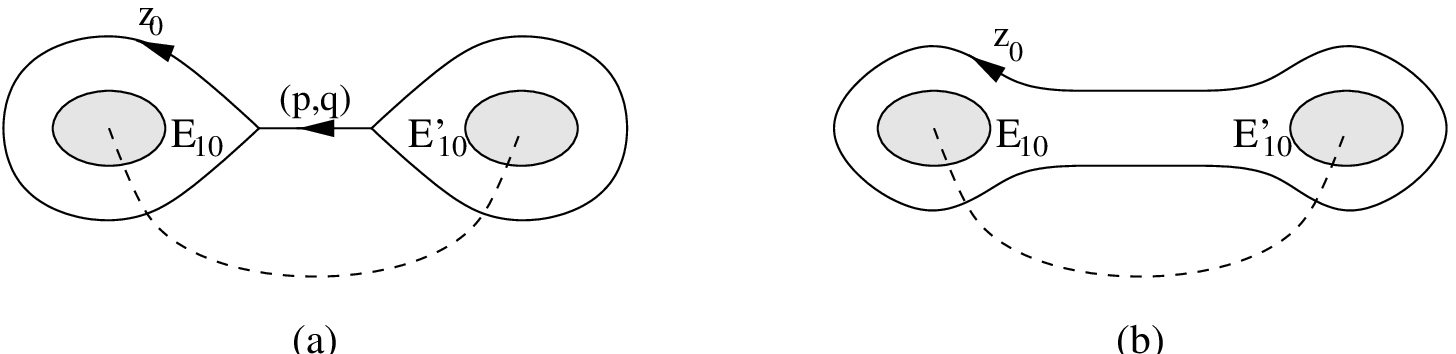}{(a) A junction which is a singlet under both
$E_{10}$'s. (b) The same singlet junction is now seen to be trivial on
a compact surface.}

The full spectrum of junctions, however is not just those that are in
the adjoint of $E_{10} \oplus E_{10}$.  Although any junction can be
expressed as $\mJ = \mJ_1 + \mJ_2$ on the two ${\bf E_{10}}$'s, the
only requirement for it to be BPS is that $\mJ^2 \geq -2$, and the
existence of a BPS junction is then guaranteed by the geometry of K3
\cite{wolfson,vsb} with a connected holomorphic surface for the
corresponding 2-cycle \cite{wolfson}.  The two pieces need not both
satisfy $\mJ^{2}_{1}\geq -2$ and $\mJ^{2}_{2}\geq -2$, and so
one of them need not correspond to a root of $E_{10}$, but to some
other weight vector.

Both the $E_9\oplus E_9$ and the $E_{10}\oplus E_{10}$ configurations
nicely capture aspects of the homology lattice of K3. The
$E_{10}\oplus E_{10}$ configuration, however, is complete since there
is a one to one correspondence between the twenty Dynkin labels of a
weight vector and the twenty invariant charges of an arbitrary
junction. On the other hand the $E_9\oplus E_9$ lattice is only
eighteen dimensional and junctions cannot be reconstructed from the
weight vectors alone. 

%%% Both the $E_9\oplus E_9$ (obtained from
%%% $\widehat{E}_9\oplus\widehat{E}_9$ after modding out by the two
%%% imaginary roots \myref{deltaequiv}) and the $E_{10}\oplus E_{10}$
%%% configurations nicely capture aspects of the homology lattice of
%%% K3. The $E_{10}\oplus E_{10}$ however, seems to be more essential
%%% in that there is a one-to-one correspondence between the Dynkin
%%% labels and the invariant charges. On the other hand the $E_9\oplus
%%% E_9$ lattice is two less dimensional and the junctions are not
%%% uniquely specified by the representation.

{\bf Other configurations.} If we consider asymmetric configurations
one can get even larger algebras.  It is well-known that no more than
18 branes can be made mutually local, corresponding to the fact that
the elliptic fibration of K3 is determined by 18 parameters
\cite{vafa}.  To see such a configuration, begin with
(\ref{enineenine}) and apply a global $\sl2z$ transformation with the
$\mA$-brane monodromy $K_A$ to convert into the ${\bf \tilde{E}_N}$
presentation:
\begin{eqnarray}
\label{threeline}
(\mA^8\, \mB\mC\mC\, \mX_{\bf [3,1]})\,\, (\mA^8\, \mB\mC\mC
\,\mX_{\bf [3,1]}) &\ra& (\mA^9 \, \mX_{\bf [2,-1]} \, \mC \, 
\mX_{\bf [4,1]}) \,  (\mA^9 \, \mX_{\bf [2,-1]} \, \mC \, 
\mX_{\bf [4,1]})  \nn \\
&=& \mA^{18} \, \mX_{\bf [2,-1]} \, \mC \, \mX_{\bf [4,1]} \, 
\mX_{\bf [2,-1]} \, \mC \, \mX_{\bf [4,1]}   \\
&=& \mA_{\bf 17} \, {\bf \widehat{E}_0 \, \widehat{E}_0} \,, \nn
\end{eqnarray}
The identification of the ${\bf A_{17}\, \widehat{E}_0\,
\widehat{E}_0}$ combination is consistent with this algebra arising in
\cite{GMS}. The middle configuration in \myref{threeline} can be
turned into:
%%% Imposing global constraints makes trivial the junctions between a
%%% $(\mX_{\bf [2,-1]} \, \mC)$ pair and the rest: 
\begin{eqnarray} 
(\mX_{\bf [4,1]}\, \mA^{18} \, \mX_{\bf [2,-1]}\, \mC\, 
\mX_{\bf [4,1]}) \,\, (\mX_{\bf [2,-1]} \mC ) \,.
\end{eqnarray}
The left-hand piece can be thought of as ``${\bf E_{18}^H}$'' , while
the right-hand piece is just ${\bf E_0}$.

%%%t If we force the branes forming $\widehat E_9$ together we seem to
%%%t decompactify the theory back to ten dimensions. On the other hand
%%%t it is not clear how to do this for the $E_{10}$ case.

\subsection{Metrics associated to 7-branes} 

Some intuition can be gained by examining the metrics on the two
sphere.  First we claim that any ${\bf D_N}$ configuration leads to a
conical geometry with deficit angle $\pi$.  This is certainly
well-known for the ${\bf D_4}$ case, and is proven as follows.  We
recall \cite{vgs} that the metric on the two sphere is of the form
$ds_{p,q} = |h_{p,q}(z) dz|$ where
\be
\label{meph}
h_{p,q} (z) = (p-q\tau) \,\eta^2 (\tau) \prod_i \, (z-z_i)^{-1/12} \, .
\ee
We consider the bunch of branes centered around $z=0$, close to each
other so that for sufficiently large $z$ we can take all $z_i$ in the
above expression to be zero.  The branch cuts go downwards from $z=0$.
The monodromy of the configuration is used to get an expression for
$\tau (z)$ and then plugged back into (\ref{meph}) to get the
$z$-behavior of $h_{p,q}$.  We will consider the metric $ds_{1,0}$ as
seen by $(1,0)$ strings.The monodromy $K$ of the ${\bf D_N}$
configuration indicated in the main table implies that $\tau_R =
\tau_L - (N-4)$, where $\tau_{R}$, $\tau_{L}$ denote the values of
$\tau$ immediately to the right and immediately to the left of the
cut, respectively. Such monodromy requires the behavior
\be
\label{howlook}
\tau(z) \sim  {1\over 2\pi i} \, (N-4) \ln z \,, \qquad -\pi/2 <
\hbox{arg}\, z < 3\pi/2\,.
\ee
Note moreover that as $z\to 0$, we find Im $\tau \to i\infty\cdot
(N-4)$ which is consistent if $N \geq 4$, but inconsistent if
$N<4$. This, of course, only means that we cannot get a genuine
singularity for $N<4$, as we have discussed repeatedly. For $\tau \to
i\infty$ we also have $\eta^2 (\tau) \sim \exp (2\pi i
\tau/12)$. Thus, for $N\geq 4$, and given that we have a total of
$(N+2)$ branes, we get
\be
\label{meteqn}
h_{1,0} (z) \sim  z^{{N-4}\over 12} \, z^{- {N+2\over 12}} \sim
z^{-1/2} \,.
\ee
This behavior indicates that away from the singularity we have a flat
conical geometry with defect angle of $\pi$, as claimed.

The computations are rather similar for the affine exceptional series
${\bf \widehat E_N}$.  The relevant monodromy $K$ indicated in the
main table implies that $\tau_R = \tau_L + (9-N)$, which requires
\be
\tau(z) \sim  -\,{1\over 2\pi i} \, (9-N) \ln z \,, \qquad -\pi/2 <
\hbox{arg}\, z < 3\pi/2\,.
\ee
Note that for the range of interest $N \leq 9$ as $|z\,|\to 0$, we
find Im $\tau \to -i\infty\,$ indicating again that the branes cannot
be brought together.  On the other hand, for $|z| \to \infty$ we get
$\tau \to +i\infty$ and the behavior of the metric far away from the
configuration can be discussed.  Since the number of branes is $(N+3)$
this time we get
\be
\label{meteqnn}
h_{1,0} (z) \sim  z^{-{(9-N)\over 12} }\, z^{- {N+3\over 12}} \sim
{1/ z} \,\,.
\ee
The metric $ds_{1,0} = |dz/z|$ away from any $\widehat {\bf E}_{\bf
N}$ configuration is that of a flat cylinder; effectively we have a
deficit angle of $2\pi$. This is in accord with the expectation of the
particular case when two nearby ${\bf D_4}$ configurations are thought
as a ${\bf \widehat E_9}$ configuration. The defect angle of $\pi$
carried by each ${\bf D_4}$ adds up to a total defect angle of $2\pi$
for the complete ${\bf \widehat E_9}$ configuration.  Two affine
configurations carry the necessary defect angle to produce a two
sphere.

For configurations leading to Lorentzian algebras the nature of the
metric is not easily elucidated. For example the ${\bf E_{10}}$
configuration has a monodromy conjugate to $\left({\,\,0\;\;1}\atop{\!
-\!1\;3}\right)$, implying that $\tau \to 1/(3-\tau)$. The fact that
the fixed points of this transformation are irrational numbers on the
real line---points that are not conjugate to points in the modular
domain--- indicates that nowhere in the sphere can the metric be
expected to be simple.  An understanding of metrics on the sphere for
the case of the double ${\bf E_{10}}$ configuration would be of
interest.

%%%%%%%%%%%%%%%%%%%%%%%%%%%%%%%%%%%%%%%%%%%%%%%%%%%%%%%%%%%%%%
\newsection{Conclusion and Open Questions} 

We presented a systematic survey of the spectrum generating algebras
arising on 7-brane configurations. Relaxing the condition that the
branes be collapsible to a singularity of K3, a wide range of algebras
were found, including semisimple, affine and indefinite Kac-Moody
algebras as well as more exotic ones. We showed that on the 24
7-branes of the full K3 a $E_{10}\oplus E_{10}$ algebra arises
naturally.

Many questions are still open. We used three $\sl2z$ invariants
(conjugacy class of the overall monodromy, number of branes and
constraints on the asymptotic charges) to classify the equivalent
configurations and thus the algebras.. These were sufficient for all
the configurations we have examined but there remains the possibility
that other invariants are necessary to uniquely identify the resulting
algebra. We also observe that not every simply laced Lie algebra is
realized, for example, only the algebras in the exceptional series
admit affine extensions.

Although the root lattice of the $E_{10}\oplus E_{10}$ algebra can be
identified with the homology of the junctions on K3, the algebra
generated by the complete set of BPS junctions is larger than
$E_{10}\oplus E_{10}$. It would be interesting to identify the algebra
for which the full set of BPS junctions is the adjoint representation.
The Narain lattice arising from compactification on a two torus of the
heterotic string is also equivalent to the root lattice of $E_{10}
\oplus E_{10}$. This being the heterotic dual to the IIB
compactification we have examined, it may be possible to relate
explicitly the junction lattice to the Narain lattice.

The algebras on the 7-branes should constrain and organize the BPS
spectrum of the corresponding D3 brane probe theories.  We believe
that the affine and indefinite algebras studied here will prove useful
in studying the spectrum of the field theories on D3 brane probes.

%\vfill
%\break

%%%%%%%%%%%%%%%%%%%%%%%%%%%%%%%%%%%%%%%%%%%%%%%%%%%%%%%%%%%%%%%%%%%%%
\subsection*{Acknowledgments}

We are grateful to V. Kac for invaluable help in identifying various
infinite dimensional algebras. Thanks are due to A.~Hanany, C.~Vafa,
E.~Witten and D.~Zagier for their useful comments and questions and to
R.~Borcherds for very helpful correspondence.

This work was supported by the U.S.~Department of Energy under
contract \#DE-FC02-94ER40818.

\section*{Appendix: Binary quadratic forms and ideal classes
of quadratic fields} 
\setcounter{equation}{0}

In this appendix we elaborate on the relation between binary quadratic
forms and ideal classes in quadratic fields and explain how to
calculate the number of conjugacy classes of binary quadratic forms of
given discriminant $\widehat{d}$.  For more details see
\cite{Buell,Hecke,Borevich}

{\bf Algebraic number theory:} A number $\theta$ is called an {\em
algebraic number} if it is a root of a polynomial with coefficients in
$\bbbq$, and it is called an {\em algebraic integer} if it is a root
of a polynomial whose leading coefficient is 1 and all the others are
integers. By adjoining an algebraic number $\theta$ to the rational
field, one obtains an extension of $\bbbq\ $, which is also a field
and is denoted by $\bbbq~(\theta)$. The set of all algebraic integers
of $\bbbq~(\theta)$ forms a ring, $\mathcal{O}(\theta)$ which has many
properties similar to those of the ring of usual integers, for example
one can define the notion of a prime number in an exactly analogous
way.

A subset $I\subset \mathcal{O}(\theta)$ is an {\em ideal} in $\bbbq~
(\theta)$ if $\alpha,\beta\in I$ and $\lambda,\mu\in
\mathcal{O}(\theta)$ implies $\lambda\alpha+\mu\beta\in I$ and it is
called a {\em prime ideal} if $\alpha,\beta\in \mathcal{O}(\theta),
\alpha \beta \in I$ implies that either $\alpha\in I$ or $\beta \in
I$. An amazing fact is that in any ring of integers every ideal can be
uniquely factorized into prime ideals. When $I$ is the set of all
multiples of some element of $\mathcal{O}(\theta)$ then we call it
{\em principal} and a ring of integers is called a {\em principal
ideal domain} if all its ideals are principal.  (The ring of ordinary
integers has this property: any ideal of $\bbbz$ is the set of all
multiples of a given integer $n$.)  Principal ideal domains have the
property that each integer can be uniquely factorized in terms of
prime integers, thus these are unique factorization domains as well.

In a generic ring the non-uniqueness of factorization is characterized
by the number of non-principal ideals as follows. One can extend the
set of ideals of $\mathcal{O}(\theta)$ to a multiplicative group,
$I(\theta)$ by adding the so called fractional ideals of $\bbbq~
(\theta)$ to it. The set of principal ideals $P(\theta)$ form a
subgroup of $I(\theta)$. The quotient group
$C(\theta)=I(\theta)/P(\theta)$ is called the {\em class group} of
field $\bbbq~ (\theta)$ and its order is the {\em class number} of the
algebraic extension. A fundamental theorem of algebraic number theory
is that for any finite algebraic extension of the rational numbers,
the class group is always a finite group.

{\bf Quadratic fields:} For  $\theta$ satisfying $\theta^{2}-D=0$ with
$D$ a square free integer, $\bbbq~ (\theta)$ is called a {\em
quadratic field}. The quadratic field is called real or imaginary when
$D>0$ or $D<0$, respectively. A generic element of $\bbbq~ (\sqrt{D})$
can be written as
\begin{equation}
\alpha = x+y\sqrt{D},~~x,y\in \bbbq\;,
\end{equation}
while its {\em conjugate} $\alpha'$, {\em norm} $N(\alpha)$ and
{\em trace} $S(\alpha)$ are defined as:
\begin{equation}
\alpha'=x-y\sqrt{D},\;\;\;\;\;
N(\alpha)=\alpha\cdot\alpha',\;\;\;\;\;S(\alpha)=\alpha+\alpha'.
\end{equation}
When $N(\alpha),S(\alpha)\in\bbbz$ then $\alpha\in
\mathcal{O}(\sqrt{\!D})$ thus the following holds for the set of
algebraic integers $\mathcal{O}(\sqrt{D})\subset\bbbq~(\sqrt{D})$: \\
$\bullet$ If $D\!=\!2,3~(mod ~4)$ then $\{1,\sqrt{D}\}$ is a
basis of $\mathcal{O}(\!\sqrt{\!D})$ with discriminant
$\widehat{d}=\left|{1\;\;\;\;\sqrt{\!D} \atop
                    1\;-\!\sqrt{\!D}}\right|^2=4D$
$\bullet$ If $D\!=\!1 (mod ~4)$ then $\{1,\frac{1+\sqrt{D}}{2}\}$
is a basis of $\mathcal{O}(\!\sqrt{\!D})$ with discriminant
$\widehat{d}=\left|{1\;\;\;\frac{1\!+\sqrt{\!D}}{2} \atop
                    1\;\;\;\frac{1\!-\sqrt{\!D}}{2}}\right|^2 = D$.
In both cases the basis of algebraic integers is given by
$\{1,\frac{\widehat{d}+\sqrt{\widehat{d}}}{2}\}$, where $\widehat{d}$
is the discriminant of the field ~$\bbbq~(\sqrt{D})$.

Two ideals $M$ and $P\subset \mathcal{O}(\sqrt{D})$ are {\em
equivalent} ($M\sim P$) if $\;\exists\;\lambda\in \bbbq~(\sqrt{\!D})$
such that $M=\lambda \cdot P$ and they are {\em strictly equivalent}
$(M\approx P)$ if $N(\lambda)>0$. For imaginary fields equivalence
implies strict equivalence because $N(\lambda)\!>\!0,\;\forall
\lambda\in\bbbq~(\sqrt{\!D})$, while in the real case $M\sim P$
implies either $M\approx P$ or $M\approx \sqrt{\widehat{d}}~P$ as
$N(1)>0$ and $N(\sqrt{D})<0$ and thus an equivalence class of ideals
splits into at most two strict equivalence classes. Let us denote with
$h(\widehat{d})$ and $h_{0}(\widehat{d})$ the number of equivalence
classes and that of strict equivalence classes, respectively, of the
quadratic field $\bbbq~(\sqrt{D})$ with discriminant $\widehat{d}$.
Then $h_{0}(\widehat{d})\leq 2h(\widehat{d})$. Moreover, equivalence
implies strict equivalence exactly when $(\sqrt{D})\approx (1)$ which
we can write as $\sqrt{D}\epsilon =\lambda$ for some unit $\epsilon$.
For $D>0$, $N(\lambda)>0$ only when $N(\epsilon)=-1$ which implies
that the fundamental unit satisfies $N(\epsilon_{F})=-1$.  Thus for a
square free integer $D$ we have the following result:
\begin{equation}
h_{0}(\widehat{d})=\left\{\begin{array}{rcccl}
h(\widehat{d}), &\mbox{if}& D<0\,, && \\
h(\widehat{d}),  &\mbox{if}& D>0 &\mbox{and}& N(\epsilon_{F})=-1\,,\\
2h(\widehat{d}), &\mbox{if}& D>0 &\mbox{and}& N(\epsilon_{F})=+1\,.
\end{array}
\right.
\end{equation}

%%%%%%%%%%%%%%%%%%%%%%%%%%%%%%%%%%%%%%%%%%%%%%%%%%%%%%%%%%%%%%%%%%%%
{\bf Ideals and binary quadratic forms:} Let $M=(\alpha,\beta)$ be an
ideal in ~$\bbbq~(\sqrt{D})$ with basis $\{\alpha,\beta\}$ then we
associate with it the following binary quadratic form:
\begin{equation}
Q_{M}(x,y)=(\alpha x+\beta y)(\alpha' x+\beta'y)/q,\mbox{~~where~~}
q=\gcd(\alpha\alpha',\alpha\beta'+\beta\alpha',\beta\beta').
\end{equation}

If $P\approx M$ and $\{\bar{\alpha},\bar{\beta}\}$ is a basis of $P$
then $\{\lambda\bar{\alpha},\lambda\bar{\beta}\}$ is a basis of $M$
for some $\lambda$ of positive norm and there exists a matrix $K$ of
determinant $+1$ such that

\begin{equation}
\pmatrix{\lambda\bar{\alpha}\cr\lambda\bar{\beta}}=
K\pmatrix{\alpha\cr\beta}.
\end{equation}
Thus the quadratic forms associated with the two ideals are related by
an $\sl2z$ transformation and thus are equivalent. The values that
this quadratic form takes are exactly the norms of the integral ideals
in the corresponding equivalence class of
ideals.

Conversely, given a quadratic form $Q(x,y)=Ax^{2}+Bxy+Cy^{2}$, we can
associate the following ideal $M(Q)$ with it,
\begin{equation}
M(Q)=(\alpha,\beta)=\left\{
\begin{array}{lcc}
(A,\frac{B+\sqrt{\widehat{d}}}{2})  &\mbox{if}& A>0 \\
(A,\frac{B+\sqrt{\widehat{d}}}{2}) \sqrt{\widehat{d}} &\mbox{if}& A<0
\end{array}
\right\}, \;\;\;\;\; \widehat{d}=B^{2}-4AC.
\end{equation}
Under a transformation by an $\sl2z$ matrix $K=\pmatrix{a&b\cr c&d}$
the quadratic form becomes $Q'(x,y)=A'x^{2}+B'xy+C'y^{2}$ where
\begin{equation}
\pmatrix{A'\cr B'\cr C'}=\pmatrix{a^{2}&ac&c^{2}\cr 2ab&ad+bc&2cd
\cr b^{2}&bd&d^{2}}\pmatrix{A\cr B\cr C}\equiv
S\pmatrix{A\cr B\cr C}.
\end{equation}
The matrix $S$ belongs to $SL(3,\bbbz)$ and $TrS=(TrK)^{2}-1$. The
basis of the ideal associated with the quadratic form $Q$ undergoes an
$\sl2z$ transformation as well as scaling by a positive norm
number. Since ideals whose basis are related by an $\sl2z$
transformation and an overall scaling are (strictly) equivalent we see
that ideals associated with equivalent quadratic forms are strictly
equivalent. If we denote by $h_{+}(\widehat{d})$ the number of
equivalence classes of binary quadratic forms with discriminant
$\widehat{d}$, then
\begin{equation}
h_{+}(\widehat{d})=h_{0}(\widehat{d}).
\end{equation}

{\bf Jacobi symbol:} Let the prime factorization of $a$ be
$a=\prod_{i}p_{i}^{c_{i}}$ and let $p$ be an odd prime,
then we define $(a/p)\equiv\prod_{i}(p_{i}/p)^{c_{i}}$ where
\begin{equation}
(p_i/p)\equiv\left \{ \begin{array}{rcl}
0, &\mbox{if}& p_i\equiv 0 ~(~mod~p)\,,
\\
+1, &\mbox{if}& x^{2}\equiv p_i ~(~mod~p) \;\mbox{is solvable}\,,
\\
-1, &\mbox{if}& x^{2}\equiv p_i ~(~mod~p) \;\mbox{is unsolvable\,,}
\end{array}
\right.
\end{equation}
while for $p=2$ we define
\be
(a/2)=\left \{
\begin{array}{rlc}
0, & \mbox{if $a$ is even\,,}
\\
+1,&\mbox{if $a\equiv 1~(mod ~8)$,}
\\
-1,& \mbox{if $a \equiv 5~(mod ~8)$.}
\end{array}
\right.%\}%
\ee

{\bf Theorem:}\cite{Cohn} If $\widehat{d}=f^{2}d>0$ is such that $d$
is the discriminant of some quadratic field then
\be
\label{mainformula}
h_{+}(f^{2}d)=h_{0}(f^{2}d)=\left \{
\begin{array}{rcl}
h(d)f\prod_{p|f}(1-\frac{(d/p)}{p})\cdot \frac{1}{E},
&\hspace{.2in}\mbox{if}&N(\epsilon_{f})=-1\,, \\ \\
2h(d)f\prod_{p|f}(1-\frac{(d/p)}{p})\cdot \frac{1}{E},
&\hspace{.2in}\mbox{if}&N(\epsilon_{f})=+1\,.
\end{array}
\right. %\}%
\ee
In the above equation $\epsilon_{f}$ is the fundamental unit of
$\mathcal{O}$$_{f}(\sqrt{D})$, the ring of integers generated by
$\{1,f(\frac{d+\sqrt{d}}{2})\}$. $E$ is the least positive integer
such that $\epsilon_{F}^{E}=\epsilon_{f}$, where $\epsilon_{F}$ is the
fundamental unit of $\mathcal{O}$$(\sqrt{D})$, the ring of integers of
~$\bbbq~(\sqrt{D})$ and $p$ is a prime divisor of $f$.

We now illustrate the use of this theorem on a concrete example. To
this end let us calculate the number of conjugacy classes in $\sl2z$
with trace $t=6$. The discriminant is $\widehat{d}=t^{2}-4=2^{5}$
which implies $f=2$ and $d=4D=8$. Thus the quadratic field we are
considering is $\bbbq~(\sqrt{2})$.  The fundamental unit of the ring
of integers $\mathcal{O}$$(\sqrt{2})$ is $\epsilon_{F}=1+\sqrt{2}$
while that of the subring $\mathcal{O}$$_{2}(\sqrt{2})$ is
$\epsilon_f=3+2\sqrt{2}=\epsilon_{F}^{2}$ which implies $E=2$ and
$N(\epsilon_{f})=+1$. The class number $h(2)$ of $\bbbq~(\sqrt{2})$ is
1 \cite{Borevich} and direct substitution into \myref{mainformula}
yields $h_{+}(2^{5})=2$. We list a few more examples in the following
table.

\begin{equation}
\begin{array}{|c|c|c|c|c|c|c|c|c|c|c|}
\hline
t & \hat{d}=df^2 &f&d&\bbbq~(\sqrt{D})
  &\epsilon_F&\epsilon_f& N(\epsilon_f\!) &
E & h(D) & h_+(\widehat{d})
\\ \hline\hline
3 & 5  &1& 5&\bbbq~(\sqrt{5})& \frac{1+\sqrt{5}}{2}
                             & \frac{1+\sqrt{5}}{2}   &-1 & 1 & 1 & 1
\\ \hline
6 & 32 &2& 8&\bbbq~(\sqrt{2})& 1\!+\!\!\sqrt{2}
                             & 3+2\sqrt{2}            &+1 & 2 & 1 & 2
\\ \hline
11& 117&3&13&\bbbq~(\sqrt{13})& \frac{3+\sqrt{13}}{2}
                             & \frac{11+3\sqrt{13}}{2}&+1 & 2 & 1 & 2
\\ \hline
14& 192&4&12&\bbbq~(\sqrt{3})& 2\!+\!\!\sqrt{3}
                             & 7+4\sqrt{3}            &+1 & 2 & 1 & 4
\\ \hline
\end{array}
\end{equation}

Table \ref{numberofclasses} lists the conjugacy classes for $-15\leq
t\leq +15$.

\begin{table}[h]
\begin{center}
\begin{tabular}{||c|c|c|c||}
\hline
$\rule{0mm}{7mm}$$t=TrK$ & $\widehat{d}=t^{2}-4$ & $H(t)$ & 
$K$ \\ \hline
\hline
$\rule{0mm}{5mm}$0    & $-4$  &  2 &  $ K=(0,1; -1,0), K^{-1}$ \\ \hline
$\rule{0mm}{5mm}$$1$    & $ -3$  &  2 & $K=(0,1; -1,1),K^{-1}$ \\
\hline
$\rule{0mm}{5mm}$$2$    &   0    & $\infty$ & $~~~~~~~K=(1 ,
-n; 0 , 1), n
\in
\bbbz$
\\ \hline
$\rule{0mm}{5mm}$$3 $ &   5  &  1 & $K=(0 , 1; -1 ,3)$ \\ \hline
$\rule{0mm}{5mm}$$4$  &  $12$  & 2 &  $K=(0,1;-1,4), K^{-1}$
\\
\hline
$\rule{0mm}{5mm}$$5$  & $21$    & 2 &
$K=(0,1; -1,5),K^{-1}$\\ \hline
$\rule{0mm}{5mm}$$6$  & $32$      & 2 &
$K=(0,1;-1,6), K^{-1}$\\ \hline
$\rule{0mm}{5mm}$$7$  & $45$    & 2 &
$K=(0,1;-1,7), K^{-1}$\\ \hline
$\rule{0mm}{5mm}$$8$  &$60$   & 4
&$K_{1}=(0,1;-1,8),K_{2}=(1,2;3,7), K_{1}^{-1},K_{2}^{-1};$\\ \hline
$\rule{0mm}{5mm}$$9$  & 77        & 2
& $K=(0,1;-1,9),K^{-1}$    \\ \hline
$\rule{0mm}{5mm}$$10$ &   $96$     & 4 &  $K_{1}=(0,1;-1,10),
K_{2}=(1,4; 2,9),K_{ 1}^{-1},K_{2}^{-1}$
\\ \hline
$\rule{0mm}{5mm}$$11$ &   $117$    &
2 &    $ K=(0,1;-1,11),K^{-1}$   \\ \hline
$\rule{0mm}{5mm}$$12$ &   $140$   & 4   & $K_{1}=(0,1;-1,12),
K_{2}=(1,2;5,11),K_{1}^{-1},K_{2}^{-1}$  \\ \hline
$\rule{0mm}{5mm}$$13$ &$165$      & 4&  $K_{1}=(0,1;-1,13),
K_{2}=(2,3;7,11),K_{1}^{-1}, K_{2}^{-1}$
\\ \hline
$\rule{0mm}{5mm}$$14$ &$192$     & 4&  $K_{1}=(0,1;-1,14),
K_{2}=(1,2;6,13),K_{1}^{-1}, K_{2}^{-1}$
\\ \hline
$\rule{0mm}{5mm}$$15$ &$221$      &
2&$ K=(0,1;-1,15),K^{-1}$\\ \hline
\end{tabular}
\end{center}
\caption{The number $H(t)$ of conjugacy classes of $\sl2z$ at a given
trace $t$, as well as the (factorized) discriminant $\widehat{d}$ of
the associated quadratic form and matrix representatives for each
class.  In the last column the matrices are denoted as 
$(a,b\,;c,d)\equiv\left({a\;b\atop c\;d}\right)$.
\label{numberofclasses}}
\end{table}

%%%%%%%%%%%%%%%%%%%%%%%%%%%%%%%%%%%%%%%%%%%%%%%%%%%%%%%%%%%%%%%%%%%%

\end{document}